\documentclass[12pt]{article}
\usepackage{graphicx}
\usepackage{amsfonts}
\usepackage{amssymb,amsmath}
\usepackage{graphicx}

\newcommand{\simgt}{\lower.5ex\hbox{$\; \buildrel > \over \sim \;$}}
\newcommand{\simlt}{\lower.5ex\hbox{$\; \buildrel < \over \sim \;$}}
\newcommand{\largesmall}{\lower.5ex\hbox{$\; \buildrel {\LARGE >} 
\over {\small <}$}}

\def\bfk{{\bf k}}
\def\bfs{{\bf s}}

\renewcommand{\thefootnote}{\#\arabic{footnote}}
\begin{document}


\begin{center}
{\Large \bf
Window effect in the power spectrum analysis of a galaxy 
redshift survey}


\vskip .45in

{Takahiro Sato{}$^a$, Gert H\"utsi{}$^b$, Gen Nakamura{}$^a$, Kazuhiro Yamamoto{}$^a$}


\vskip .45in
{
{}$^a$Graduate School of Physical Sciences, Hiroshima University,
Higashi-Hiroshima 739-8526,~Japan\\
{}$^b$
Tartu Observatory, EE-61602 T\~{o}revere, Estonia}

\vskip .45in


\begin{abstract}
We investigate the effect of the window function on the multipole power 
spectrum in two different ways. First, we consider the convolved power 
spectrum including the window effect, which is obtained by following
the familiar (FKP) method developed by Feldman, Kaiser and Peacock.
We show how the convolved multipole power spectrum is related 
to the original power spectrum, using the multipole 
moments of the window function. Second, we investigate
the deconvolved power spectrum, which is obtained by using the Fourier
deconvolution theorem. In the second approach, 
we measure the multipole power spectrum deconvolved from the window effect. 
We demonstrate how to deal with the window effect in these two approaches, 
applying them to
the Sloan Digital Sky Survey (SDSS) luminous red galaxy (LRG) sample.
\end{abstract}



\end{center}

\section{Introduction}
One of the most fundamental problems in cosmology is the origin of
an accelerated expansion of the Universe \cite{Perlmutter,Riess}. 
A hypothetical energy component, dark energy, may explain 
the accelerated expansion \cite{PR}. 
Modification of the gravity theory is an alternative way to explain 
it. In either case, this problem seems to be deeply rooted in the 
nature of fundamental physics, which has attracted many researchers.  
Dark energy surveys which aim at measuring redshifts of huge number 
of galaxies are in progress or planned \cite{DETF,Peacock}. 
These surveys provide us with a 
chance to test the hypothetical dark energy, as well as the gravity 
theory on the scales of cosmology. 
A key for distinguishing between the dark energy and modified gravity 
theory is a measurement of the evolution of cosmological perturbations. 

Galaxy redshift surveys provide promising ways of measuring the 
dark energy properties. 
Here, a measurement of the baryon acoustic oscillations 
in the galaxy distribution plays a key role. Also, the spatial 
distribution of galaxies is distorted due to the peculiar motions, 
which is called the redshift-space distortion. 
The Kaiser effect is the redshift-space distortion in the linear 
regime of the density perturbations. It is caused by the bulk 
motion of galaxies \cite{Kaiser}.
The measurement of the Kaiser effect is thought to
be useful for testing the general relativity and other modified
gravity theories \cite{Linder,Guzzo,Reyes}. 
In these analyses, measuring the multiple power spectrum in the 
distribution of galaxies plays a key role (cf. \cite{Okumura,Cabre}).

The multipole power spectrum is useful for measuring the redshift-space
distortion \cite{CFW,Hamilton,OutramI,OutramII,YNKBN,Blake,Taruya}.
The usefulness of the quadrupole power spectrum to 
constrain modified gravity models is demonstrated 
in Refs.~\cite{Yamamoto08,Yamamoto10}, 
as well as the dark energy model \cite{YBN}. 
An estimator of the quadrupole power spectrum is developed in 
Ref. \cite{YNKBN}. 
However, the disadvantage of
the method is not being compatible with the use of the fast Fourier 
transform (FFT). In the present paper, we consider different estimators 
of the quadrupole power spectrum which allows the use of the FFT.
In this method, a full sample of a wide survey area is divided 
into smaller subsamples with equal areas. This approach was taken in 
Refs.~\cite{OutramI,OutramII}.
In this case, the effect of the window function is crucial 
as we will show in the present paper. Thus, it must be 
properly taken into account when comparing the observational 
data with theoretical predictions. 

The {\it convolved} power spectrum includes the effect of the window 
function \cite{Percival10a,Reid,Percival07a,Cole}. 
In the first half of the present paper, we consider the 
convolved power spectrum. We develop a theoretical formula 
to incorporate the window effect into the multipole power 
spectra for the first time. 
We apply this formula to the Sloan Digital Sky Survey (SDSS) 
luminous red galaxy (LRG) sample from the data release (DR) 7, 
and investigate the behavior of the window function and its
effect on the monopole and quadrupole spectra.
We demonstrate how the window effect modifies the monopole
spectrum and the quadrupole spectrum.
In the second half, we consider the {\it deconvolved} power 
spectrum, which is developed in Ref. \cite{SATO}, 
and compare it with the results of the first approach.

This paper is organized as follows:  In section 2, we briefly
review the power spectrum analysis and the window effect, 
where the convolved power spectrum is introduced. 
In section 3, using the multipole moments of the window 
function, we derive the main formula to describe how 
the convolved multipole power spectrum is related to the
original power spectrum. 
Then, a method to measure the multipole moments of the
window function is presented. We also
apply the method to the SDSS LRG DR 7. In section 4, the method
for measuring the deconvolved power spectrum is reviewed. 
Then, a comparison of the two approaches is given.
Section 5 is devoted to summary and conclusions.
In the appendix, we give a brief review of a theoretical model,
which we adopted.
Throughout this paper, we use units in which the velocity of 
light equals 1, and adopt the Hubble parameter $H_0=100h$km/s/Mpc
with $h=0.7$.

\section{Basic formulas of the FKP method}
Let us first summarize the power spectrum analysis developed by
Feldman, Kaiser and Peacock (\cite{FKP94}, hereafter FKP). 
With this formulation we 
obtain the convolved power spectrum, including the window effect.
We denote the number density field of galaxies by $n_{\rm g}({\bfs})$, 
where $\bfs=s(z)\hat\bfs$ is the three-dimensional coordinate in the
(fiducial) redshift space, $\hat \bfs$ is the unit directional vector, 
and $s(z)$ is the comoving distance of a fiducial cosmological model.
According to Ref. \cite{FKP94}, we introduce the fluctuation field
\begin{eqnarray}
F(\bfs)=n_{\rm g}(\bfs)-\alpha n_{\rm s}(\bfs),
\end{eqnarray}
where $n_{\rm g}(\bfs)=\sum_i\delta(\bfs-\bfs_i)$, with $\bfs_i$ 
being the location 
of the $i$th object; similarly, $n_{\rm s}(\bfs)$ is the density 
of a synthetic catalog that has a mean number density 
$1/\alpha$ times that of the galaxy catalog. In the present paper, 
we adopt $\alpha=0.01$.
The synthetic catalog is a set of random points without any correlation,
which can be constructed through a random process by mimicking 
the selection function of the galaxy catalog.
For $n_{\rm g}(\bfs)$ and $n_{\rm s}(\bfs)$, we assume
\begin{eqnarray}
&&\left<n_{\rm g}(\bfs_1)n_{\rm g}(\bfs_2)\right>
=\bar n(\bfs_1)\bar n(\bfs_2)[1+\xi(\bfs_1,\bfs_2)]
+\bar n(\bfs_1)\delta(\bfs_1-\bfs_2),
\\
&&\left<n_{\rm s}(\bfs_1)n_{\rm s}(\bfs_2)\right>
=\alpha^{-2}\bar n(\bfs_1)\bar n(\bfs_2)
+\alpha^{-1}\bar n(\bfs_1)\delta(\bfs_1-\bfs_2),
\\
&&\left<n_{\rm g}(\bfs_1)n_{\rm s}(\bfs_2)\right>
=\alpha^{-1}\bar n(\bfs_1)\bar n(\bfs_2),
\end{eqnarray}
where $\bar n(\bfs)$ denotes the mean number density of the galaxies,
and $\xi(\bfs_1,\bfs_2)$ is the two-point correlation function.
These relations lead to
\begin{eqnarray}
&&\left<F(\bfs_1)F(\bfs_2)\right>
=\bar n(\bfs_1)\bar n(\bfs_2)
\xi(\bfs_1,\bfs_2)
+(1+\alpha)\bar n(\bfs_1)\delta(\bfs_1-\bfs_2).
\end{eqnarray}

We introduce the Fourier coefficient of $F(\bfs)$ by
\begin{eqnarray}
&&\hspace{-1cm}
{\cal F}_0(\bfk)={\int d^3s \psi(\bfs,\bfk)F(\bfs) e^{i\bfk\cdot\bfs}
\over
[\int d^3s \bar n^2(\bfs) \psi^2(\bfs,\bfk)]^{1/2}},
\end{eqnarray}
where $\psi(\bfs,\bfk)$ is the weight function (Throughout this paper,
we assume $\psi=1$). 
The expectation value of $|{\cal F}_0(\bfk)|^2$ is 
\begin{eqnarray}
&&\hspace{-1cm}\left<|{\cal F}_0(\bfk)|^2\right>
={1\over (2\pi)^3} \int d^3k' P(\bfk')
W(\bfk-\bfk')
+(1+\alpha)S_0(\bfk)
\end{eqnarray}
with
\begin{eqnarray}
&&\hspace{-1cm}W(\bfk-\bfk')
={\big|\int d^3s \bar n(\bfs) \psi(\bfs,\bfk) e^{i\bfs\cdot(\bfk-\bfk')}\big|^2
\over \int d^3s \bar n^2(\bfs) \psi^2(\bfs,\bfk)}
\label{Wkk}
\end{eqnarray}
and 
\begin{eqnarray}
&&\hspace{-1cm}
S_0(\bfk)=
{\int d^3s \bar n(\bfs) \psi^2(\bfs,\bfk)\over\int d^3s \bar n^2(\bfs) \psi^2(\bfs,\bfk)},
\end{eqnarray}
where we used 
\begin{eqnarray}
&&\hspace{-1cm}\xi(\bfs_1,\bfs_2)
={1\over(2\pi)^3}\int d^3k P(\bfk)e^{-i\bfk\cdot(\bfs_1-\bfs_2)}.
\end{eqnarray}
Here, $W(\bfk)$ is the window function and $S_0(\bfk)$ is the shotnoise. 
The estimator of the {\it convolved} power spectrum is taken 
\begin{eqnarray}
  &&\hspace{-1cm}
P^{\rm conv}(\bfk)
=|{\cal F}_0(\bfk)|^2-(1+\alpha)S_0(\bfk),
\label{estconvP}
\end{eqnarray}
whose expectation value is
\begin{eqnarray}
&&\hspace{-1cm} \left<P^{\rm conv}(\bfk)\right>=
{1\over (2\pi)^3} \int d^3k' P(\bfk')
W(\bfk-\bfk').
\label{FkFk}
\end{eqnarray}
Hereafter, we omit $\left<\cdot\right>$, for simplicity.

\begin{figure}[h]
\begin{center}
    \includegraphics[width=0.35\textwidth]{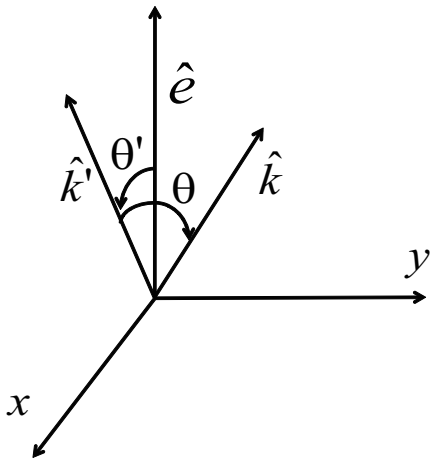}
\end{center}
\caption{A sketch of the configuration of the vectors and 
coordinate variables. 
\label{fig:angles2}}
\end{figure}
\begin{figure}[t]
\begin{center}
    \includegraphics[width=0.5\textwidth]{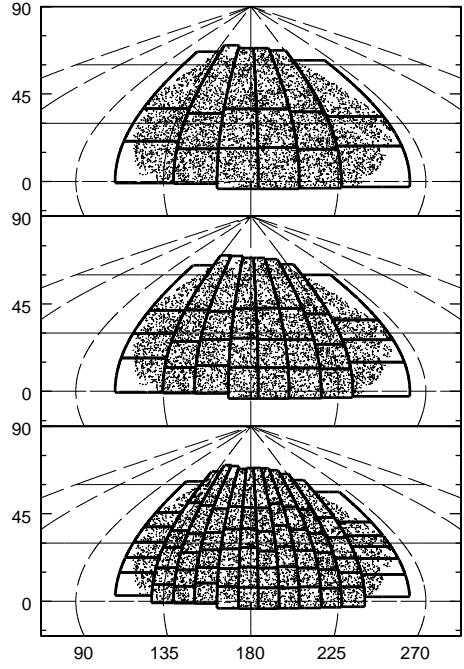}
\end{center}
\caption{Angular distribution of the SDSS LRG sample. 
In the present paper, we consider the three cases of 
the division of the full sample into subsamples. 
This figure shows the three cases of the division, where 
the full sample is divided into $18$ subsamples (upper panel), 
$32$ subsamples (middle panel)
and $72$ subsamples (lower panel), with mean area of
$397$ square degrees, $223$ square degrees and $99$ square degrees per patch, 
respectively.
\label{fig:Map}}
\end{figure}

\section{Convolved power spectrum}
In this section, using the multipole moments of the window function,
we drive the main formulas for the convolved 
multipole power spectrum, Equations (\ref{convP0f}) and (\ref{convP2f}), 
which describe the relations between 
the convolved multipole power spectrum and the original multipole 
spectrum. 
We exemplify the behavior of the multipole moments of the window 
function and the convolved spectra, using the SDSS LRG sample from
the DR 7.
\subsection{formulation}
The estimator of the monopole power spectrum should be taken as 
\begin{eqnarray}
&&\hspace{-1cm}
P_0^{\rm conv}(k)={1\over V_{k}} \int_{V_k} d^3k P^{\rm conv}(\bfk)
={1\over V_{k}} \int_{V_k} d^3k 
\bigl( |{\cal F}_0(\bfk)|^2
-(1+\alpha)S_0(\bfk)\bigr),
\end{eqnarray}
where $V_k$ is the volume of the shell in the $\bfk$-space. 
Similarly, a higher multipole power spectrum can be obtained 
\cite{YNKBN}.
Using the quantity
\begin{eqnarray}
&&\hspace{-1cm}{\cal F}_\ell(\bfk)=
{\int d^3s \psi(\bfs,\bfk)F(\bfs) e^{i\bfk\cdot\bfs}
{\cal L}_\ell(\hat\bfs\cdot\hat\bfk)
\over
[\int d^3s \bar n^2(\bfs) \psi^2(\bfs,\bfk)]^{1/2}},
\end{eqnarray}
where ${\cal L}_\ell(\mu)$ is the Legendre polynomial, and $\hat \bfk$
is the unit wavenumber vector $\hat \bfk=\bfk/|\bfk|$,
the estimator for the higher multipole power spectrum should be taken as 
(cf. \cite{YNKBN})
\begin{eqnarray}
&&\hspace{-1cm}
P_\ell^{\rm conv}(k)={1\over V_{k}} \int_{V_k} d^3k 
\bigl( {\cal F}_0(\bfk){\cal F}^*_\ell(\bfk)
-(1+\alpha)S_\ell(\bfk)\bigr),
\label{Pelll}
\end{eqnarray}
with
\begin{eqnarray}
&&\hspace{-1cm}S_\ell(\bfk)=
{\int d^3s \bar n(\bfs) \psi^2(\bfs,\bfk) {\cal L}_\ell(\hat\bfs\cdot\hat\bfk)
\over\int d^3s \bar n^2(\bfs) \psi^2(\bfs,\bfk)}.
\end{eqnarray}
The expectation value of Equation (\ref{Pelll}) is 
\begin{eqnarray}
&&\hspace{-1cm}P_\ell^{\rm conv}(k)={1\over V_{k}} \int_{V_k} d^3k
{1\over (2\pi)^3}
\int d^3 k' P({\bf k}') {\cal W}_\ell(\bfk-\bfk'),\label{233915_29Mar10}
\end{eqnarray}
where we defined
\begin{eqnarray}
&&\hspace{0cm}{\cal W}_\ell(\bfk-\bfk')=[\int d^3s \bar n^2(\bfs) \psi^2(\bfs,\bfk)]^{-1}
\nonumber
\\
&&\hspace{.5cm}\times
\int d^3s_1 \bar n(\bfs_1) \psi(\bfs_1,\bfk) e^{i\bfs_1\cdot(\bfk-\bfk')}
\int d^3s_2 \bar n(\bfs_2) \psi(\bfs_2,\bfk) e^{-i\bfs_2\cdot(\bfk-\bfk')}
{\cal L}_{\ell}(\hat \bfs_2\cdot\hat\bfk).
\end{eqnarray}
By adopting the {distant observer approximation}, we have
\begin{eqnarray}
&&\hspace{-1cm}{\cal W}_\ell(\bfk-\bfk')\simeq
W({\bf k}-{\bf k}')
{\cal L}_\ell(\hat{\bf e}\cdot \hat{\bf k}),
\end{eqnarray}
and
\begin{eqnarray}
&&\hspace{-1cm}P_\ell^{\rm conv}(k)={1\over V_{k}} \int_{V_k} d^3k
{1\over (2\pi)^3}\int d^3 k' 
P({\bf k}') W({\bf k}-{\bf k}')
{\cal L}_\ell(\hat{\bf e}\cdot \hat{\bf k}),
\end{eqnarray}
where $\hat{\bf e}$ is the unit vector along the line of sight.
We consider the shell in the Fourier space whose outer (inner) radius 
is $k_{\rm max}~(k_{\rm min})$. The volume of the shell is
$V_k\simeq 4\pi k^2\Delta k$, where
$k=(k_{\rm max}+k_{\rm min})/2$ and $\Delta k=k_{\rm max}-k_{\rm min}$, then
\begin{eqnarray}
&&\hspace{-1cm}
P^{\rm conv}_\ell(k)
={1\over 4\pi k^2\Delta k}\int_{k_{\rm min}}^{k_{\rm max}} dk k^2 
\int d\Omega_{\hat\bfk} 
{1\over (2\pi)^3}
\int d^3 {\bf k}' P({\bf k}') W({\bf k}-{\bf k}')
{\cal L}_\ell(\hat{\bf e}\cdot \hat{\bf k}).
\end{eqnarray}
Let us consider the limit $\Delta k\rightarrow 0$, then we have
\footnote{
Note that our definition of the multipole spectrum 
$P_\ell(k)$ is different from the conventional one by the factor 
$2\ell+1$ \cite{CFW,Hamilton}.}
\begin{eqnarray}
&&\hspace{-1cm}
P^{\rm conv}_\ell(k)={1\over 4\pi}
\int d\Omega_{\hat\bfk} {1\over (2\pi)^3}
\int d^3 \bfk' 
P({\bf k}') W({\bf k}-{\bf k}')
{\cal L}_\ell(\hat{\bf e}\cdot \hat{\bf k}).
\label{PPWL}
\end{eqnarray}

Now we introduce the coordinate variables to describe 
${\bf k}$ and ${\bf k}'$. For $\hat{\bf e}$ and $\hat{\bf k}'$,
we adopt
\begin{eqnarray}
&&\hspace{-1cm}
\hat{\bf e}=\left(
\begin{array}{c}
0\\
0\\
1\\
\end{array}
\right),~~
{\bf k}'=k'\left(
\begin{array}{c}
\sin\theta'\cos\varphi' \\
\sin\theta'\sin\varphi' \\
\cos\theta'
\end{array}
\right),
\end{eqnarray}
respectively.
As we consider the power spectrum and the window function averaged
over the longitudinal variable around the axis of the direction
$\hat{\bf e}$, we may choose $\hat{\bf k}'$ 
so that $\varphi'=0$ without loss of generality.
Then, we choose the coordinate variable to describe ${\bf k}$ as
\begin{eqnarray}
&&\hspace{-1cm}
{\bf k}=k\left(
\begin{array}{ccc}
\cos\theta' & 0 & \sin\theta' \\
0 & 1 & 0 \\
-\sin\theta' & 0 & \cos\theta'
\end{array}
\right)
\left(
\begin{array}{c}
\sin\theta \cos\varphi\\
\sin\theta \sin\varphi\\
\cos\theta
\end{array}
\right),
\label{righh}
\end{eqnarray}
where $\theta$ and $\varphi$ are the angle coordinates
around $\hat{\bf k}'$ so as to be the polar axis.
The matrix of the right hand side of Equation (\ref{righh}) denotes the
rotation around the $y$-axis. See figure 1 for the configuration.
Note that 
\begin{eqnarray}
&&\hspace{-1cm}
\hat{\bf e}\cdot \hat{\bf k}=
-\sin\theta'\sin\theta \cos\varphi
+\cos\theta'\cos\theta,
\label{ecka}
\\
&&\hspace{-1cm}
\hat{\bf e}\cdot \hat{\bf k}'=\cos\theta',
\label{eckb}
\\
&&\hspace{-1cm}
|{\bf k}-{\bf k}'|=\sqrt{k^2+k'^2-2kk'\cos\theta}.
\label{eckc}
\end{eqnarray}
Assuming the following formula within the distant observer approximation,
\begin{eqnarray}
&&\hspace{-1cm}
W({\bf k})=W(k,\hat {\bf e}\cdot\hat \bfk)
=\sum_{m=0,2,\cdots} W_m(k){\cal L}_{m}(\hat{\bf e}\cdot \hat{\bf k})(2m+1),
\label{Wkl}
\end{eqnarray}
and $P({\bf k}')=P({k}',\hat{\bf e}\cdot\hat\bfk')$,
Equation (\ref{PPWL}) yields
\begin{eqnarray}
&&\hspace{-1cm}P^{\rm conv}_\ell(k)=
{1\over (2\pi)^3}\int d k'dk'^2
{1\over 4\pi} \int d\Omega_{\hat\bfk}\int d\Omega_{\hat\bfk'}  
P\left(k',\hat{\bf e}\cdot \hat{\bf k'}\right)
\nonumber
\\
&&\hspace{0.5cm}
\sum_{m}W_{m}(|{\bf k}-{\bf k}'|)
{\cal L}_{m}\left( {\hat{\bf e}\cdot({\bf k}-{\bf k}')\over|{\bf k}-{\bf k}'|}
\right)
(2m+1) 
{\cal L}_\ell(\hat{\bf e}\cdot \hat{\bf k}).
\label{expressionofP}
\end{eqnarray}
Using (\ref{ecka}), (\ref{eckb}), and (\ref{eckc}), 
we can write Equation (\ref{expressionofP}) as
\begin{eqnarray}
  &&\hspace{-0.5cm}
P_0^{\rm conv}(k)= {1\over(2\pi)^2} \int_0^\infty dk'k'^2
\int_{-1}^1d\cos\theta  \int_{-1}^1 d\mu'P(k',\mu')
\bigg[
  {1\over 2}W_0(S)
\nonumber\\
  &&\hspace{0.5cm}
+{5\over 16} {W_2(S)\over S^2} 
\bigg(
(k^2+4k'^2-8 kk' \cos\theta 
+3k^2 \cos2\theta)
({3}\mu'^2-1)
\bigg)
\bigg],
\\
&&\hspace{-0.5cm}
P_2^{\rm conv}(k)= {1\over(2\pi)^2} \int_0^\infty dk'k'^2 
\int_{-1}^1d\cos\theta\int_{-1}^1 d\mu' P(k',\mu') 
\bigg[
  {1\over 16}W_0(S)(3\cos2\theta+1)
  (3\mu'^2-1)
\nonumber
\\
&&\hspace{0.5cm}
+{5\over 512} {W_2(S)\over S^2}\bigg\{
49k^2
+16k'^2
-80 kk'\cos\theta-12 k^2\cos2\theta
+48 k'^2 \cos2\theta
-48 kk' \cos3\theta 
\nonumber
\\
&&\hspace{0.5cm}
+27 k^2 \cos4\theta
+(-42 k^2 -96 k'^2 
+192 kk' \cos\theta 
-72 \cos2\theta
-288 k'^2 \cos2\theta
\nonumber
\\
&&\hspace{0.5cm}
+576kk'\cos 3\theta
-270 k^2 \cos4 \theta)\mu'^2
+(81k^2+144 k'^2 
-432 kk'\cos\theta 
+180 k^2 \cos2\theta
\nonumber\\
&&\hspace{0.5cm}
+432 k'^2\cos2\theta
-720 kk'\cos3\theta 
+315 k^2 \cos4\theta)\mu'^4
\bigg\}
\bigg],
\end{eqnarray}
where $S=\sqrt{k^2+k'^2-2kk'\cos\theta}$.
Using the relation
\begin{eqnarray}
&&\hspace{-1cm}
P({\bf k}')=P(k',\hat \bfs\cdot\hat \bfk')
=\sum_\ell P_\ell(k'){\cal L}_\ell(\hat{\bf e}\cdot \hat{\bf k}')(2\ell+1),
\end{eqnarray}
we obtain
\begin{eqnarray}
&&\hspace{-1cm}
P_0^{\rm conv}(k)={1\over(2\pi)^2}\int dk'k'^2
\bigg[
P_0(k')\int_{-1}^1d\cos\theta W_0(S)
\nonumber\\
&&\hspace{0.5cm}
+P_2(k')\int_{-1}^1d\cos\theta {5\over4}{W_2(S)\over S^2}
(k^2+4k'^2-8kk'\cos\theta
+3k^2\cos2\theta)
\bigg],
\label{convP0f}
\end{eqnarray}
\begin{eqnarray}
&&\hspace{-1cm}P_2^{\rm conv}(k)={1\over(2\pi)^2}\int dk'k'^2
\bigg[P_0(k')\int_{-1}^1d\cos\theta {1\over4}{W_2(S)\over S^2}(4k^2
+k'^2-8kk'\cos\theta
\nonumber
\\
&&\hspace{0.5cm}
+3k'^2\cos2\theta)
+P_2(k')\int_{-1}^1d\cos\theta 
\bigg({1\over 4}W_0(S)
(3\cos2\theta+1)
\nonumber\\
&&\hspace{0.5cm}
+{5\over 28}{W_2(S)\over S^2}
(-13kk'\cos\theta
+2(k^2+k'^2)(3\cos2\theta+1)
-3kk'\cos3\theta)\bigg)
\nonumber\\
&&\hspace{0.5cm}
+P_4(k')\int_{-1}^1d\cos\theta {9\over 224}{W_2(S)\over S^2}
(9k^2+16k'^2-48kk'\cos\theta
\nonumber\\
&&\hspace{0.5cm}
+4(5k^2+12k'^2)\cos2\theta
-80kk'\cos3\theta+35k^2\cos4\theta)
\bigg].
\label{convP2f}
\end{eqnarray}
%
These formulas describe how the convolved spectra, $P^{\rm conv}_0(k)$
and $P^{\rm conv}_2(k)$, are modified due to the window
effect, compared with the original spectrum. 
Using Equations (\ref{convP0f}) and (\ref{convP2f}),
we define the quantity,
\begin{eqnarray}
&&\hspace{-0.5cm}
A_\ell(k)={P^{\rm conv}_\ell(k)\over P_\ell(k)},
\label{Aell}
\end{eqnarray}
which is the correction factor connecting the original spectrum
and the convolved power spectrum.


\subsection{Measurement of the multipole moments of the window function}
In this subsection, we explain a method to measure the multipole moment
of the window function. The window function can be evaluated using the 
random catalog in a similar way of evaluating the power spectrum. 
Similar to the case of the power spectrum, we need to subtract
the shotnoise contribution. Then, we adopt the following estimator 
for the window function $W(\bfk)$, corresponding to the right hand 
side of Equation (\ref{Wkk}), 
\begin{eqnarray}
\nonumber\\
&&\hspace{-1cm}
W(\bfk)={\big|\int d^3s \alpha n_{\rm s} (\bfs) \psi(\bfs,\bfk)
e^{i\bfs\cdot\bfk}\big|^2
\over 
\int d^3s \bar n^2(\bfs)\psi^2(\bfs,\bfk)}
-\alpha S_0(\bfk).
\label{estW}
\end{eqnarray}
We consider the window function expanded  in the form of Equation (\ref{Wkl}). 
Mimicking the method to obtain the multipole 
power spectrum, we introduce
\begin{eqnarray}
&&\hspace{-1cm}
{\cal N}_\ell(\bfk)=
{\int d^3s \psi(\bfs,\bfk)\alpha n_{\rm s}(\bfs) 
e^{i\bfk\cdot\bfs}{\cal L}_\ell(\hat\bfs\cdot\hat\bfk)
\over
[\int d^3s \bar n^2(\bfs) \psi^2(\bfs,\bfk)]^{1/2}},
\end{eqnarray}
and use the following estimator for the multipole moment of the
window function, 
\begin{eqnarray}
&&W_\ell(k)={1\over V_{k}} \int_{V_k} d^3k 
\Bigl( {\cal N}_0(\bfk){\cal N}^*_\ell(\bfk)
-\alpha S_\ell(\bfk)\Bigr). 
\end{eqnarray}

\begin{figure}[thb]
\begin{center}
    \includegraphics[width=0.72\textwidth]{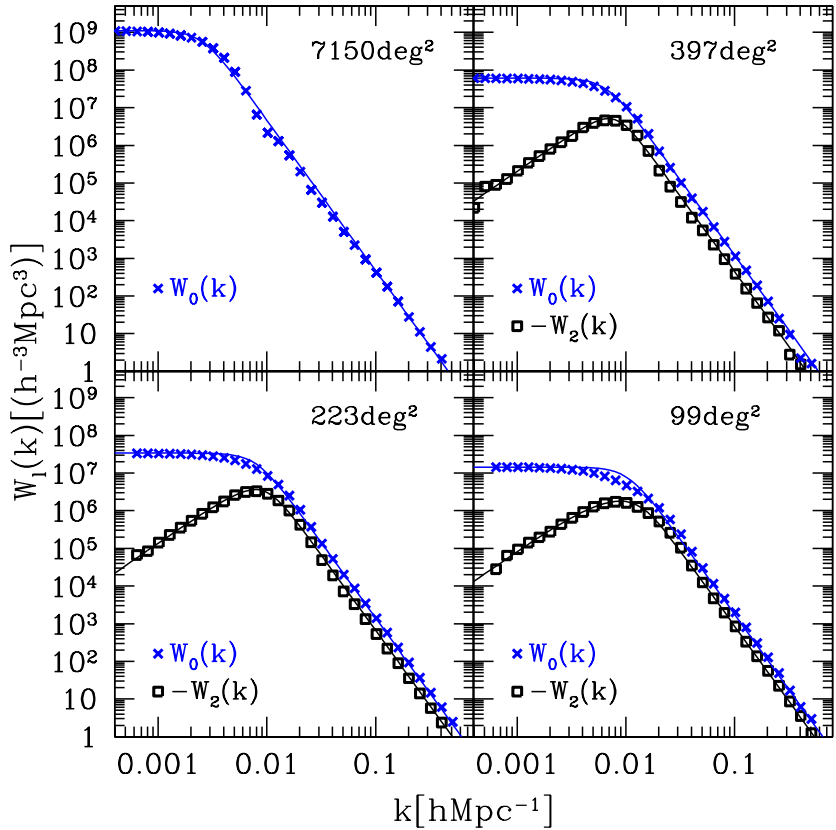}
\end{center}
\caption{$W_0(k)$ and $-W_2(k)$ as a function of $k$.
The top left panel corresponds to the full sample
without division into subsamples. In this case 
$W_0(k)$ can be measured properly, but $W_2(k)$ can not.
The other three panels represent the cases where the full sample is
divided into $18$, $32$, and $72$ subsamples, with mean area 
$397$, $223$, and $99$ square degrees per patch, respectively. 
The curves are the analytic functions, Equations (\ref{fitW0})
and (\ref{fitW2}) for $W_0(k)$ and $|W_2(k)|$, respectively. 
\label{fig:W0W2}}
\end{figure}


In the present work, we use the SDSS public data from the DR7 
\cite{DR7}.
Our LRG sample is restricted to the redshift range $z=0.16$ - $0.47$. 
In order to reduce the sidelobes of the survey window we 
remove some noncontiguous parts of the sample, which leads
us to 7150 deg${}^2$ sky coverage with the total number $N=100157$ LRGs.
The data reduction is the same as that described in 
Refs.~\cite{Hutsi,HutsiII,Yamamoto08,Yamamoto10}.
In this subsection, we show general features of the window function 
of the LRG sample. In our approach, division of the full 
sample into subsamples is necessary because the line of
sight direction is approximated by one direction $\hat{\bf e}$, 
and the distant observer approximation is required.
Each subsample is distributed in a narrow area.
We consider the three cases of the division, which are 
demonstrated in Fig.~\ref{fig:Map}. The full sample is divided 
into $18$, $32$, and $72$ subsamples, respectively. 
In those divisions of the full sample, each subsample has 
almost the same survey area, $398$, $223$, and $99$ square degrees, 
respectively. Figure \ref{fig:Map} shows the cases divided into
$18$ subsamples, $32$ subsamples and $72$ subsamples. 
Figure \ref{fig:W0W2} shows $W_0(k)$ and $W_2(k)$ as a function of 
$k$, which are obtained by averaging the results over all subsamples.
As demonstrated in Fig.~\ref{fig:W0W2}, $W_0(k)$ and $W_2(k)$ 
can be fitted in the form, 
\begin{eqnarray}
&&W_0(k)={a\over 1+(k/b)^4},
\label{fitW0}
\\
&&W_2(k)={c\over 1+(d/k)^2+(k/e)^4},
\label{fitW2}
\end{eqnarray}
where the best fitting parameters $a$, $b$, $c$, $d$ and $e$,
which depend on the division of the full sample,
are given in Table I.
%
%
\begin{table}[h]
  \begin{tabular}{l|ccccc}
   \hline
   area&$a[(h^{-1}{\rm Mpc})^3]$&$b[h{\rm Mpc^{-1}}]$&$c[(h^{-1}{\rm Mpc})^3]$&$d[h{\rm Mpc^{-1}}]$&$e[h{\rm Mpc^{-1}}]$\\
   \hline\hline
   ${\rm 7150deg^2}$&$1.111\times10^9$&0.002546&&&\\
   \hline
   ${\rm 397deg^2}$&$6.060\times10^7$&0.006680&$-2.5\times10^7$&0.011&0.0065\\
   \hline
   ${\rm 223deg^2}$&$3.382\times10^7$&0.008033&$-1.0\times10^7$&0.0085&0.009\\
   \hline
   ${\rm 99deg^2}$&$1.440\times10^7$&0.01050&$-3.0\times10^6$&0.006&0.013\\
   \hline
  \end{tabular}
  \caption{Values of the best fitting parameters for $W_0(k)$ and $W_2(k)$
  in Eqs.~(\ref{fitW0}) and (\ref{fitW2}), respectively.}
\end{table}

\begin{figure}[h]
\begin{center}
\begin{tabular}{cc}
    \includegraphics[width=0.45\textwidth]{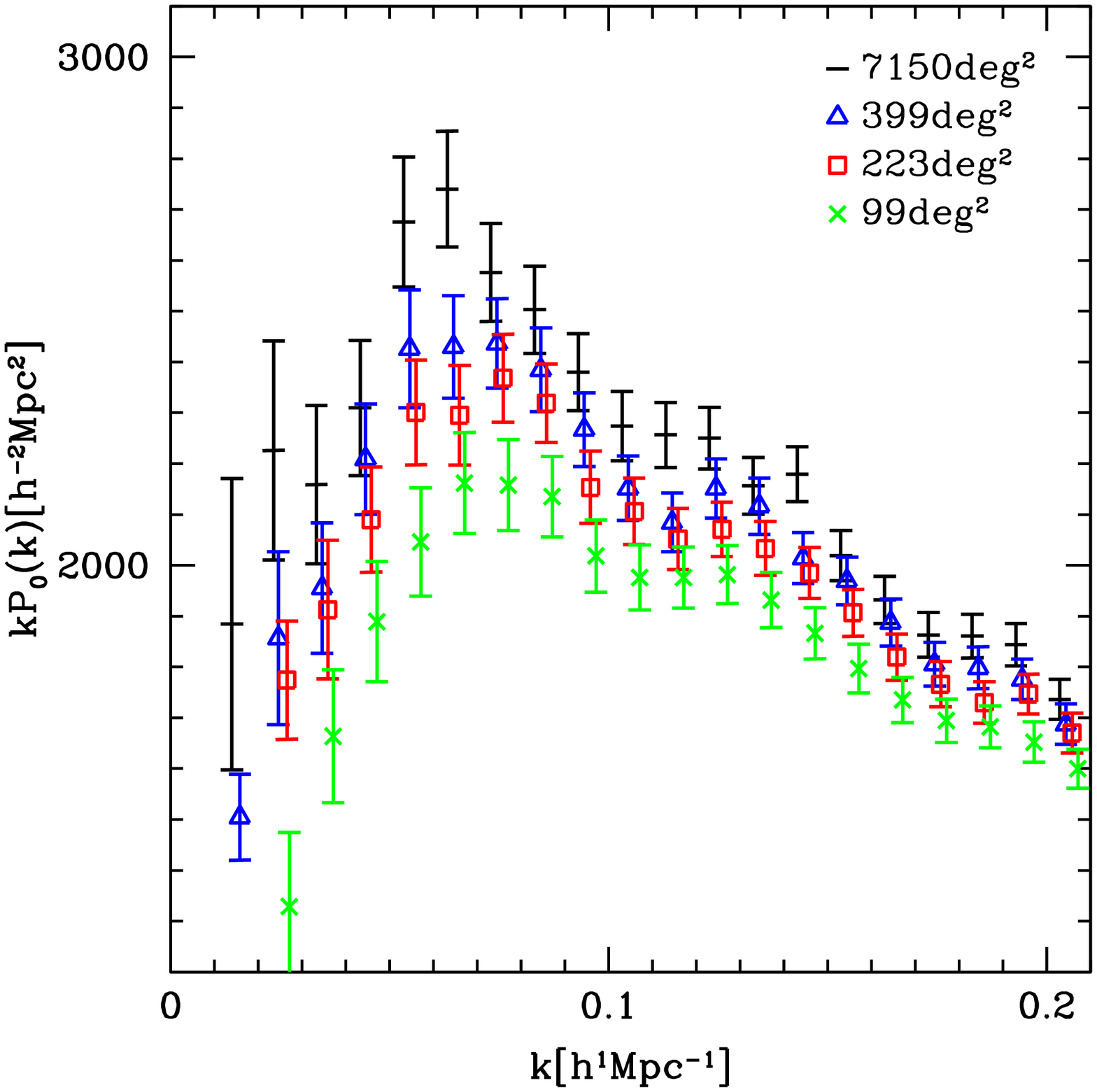}
&
   \includegraphics[width=0.45\textwidth]{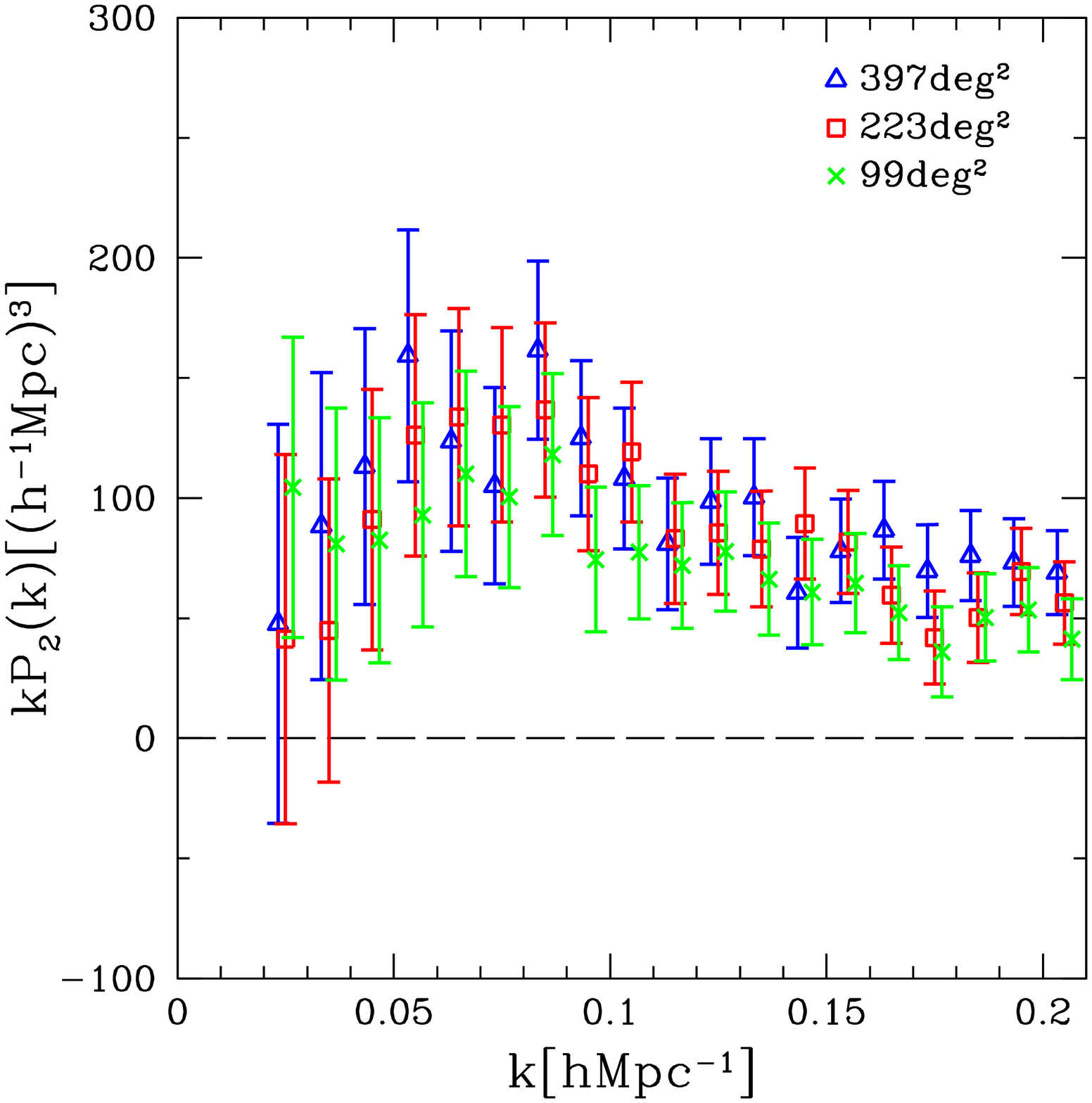}
\end{tabular}
\end{center}
\caption{Convolved monopole power spectrum multiplied by the wavenumber 
$kP^{\rm conv}_0(k)$ (left panel) 
and the quadrupole spectrum $kP^{\rm conv}_2(k)$ (right panel), respectively.
In the left panel the curves from top to bottom correspond to the cases with 
no division of the full sample, to the division into $18$, 
into $32$, and into $72$ subsamples,
respectively. In the right panel, the results are for the cases with the 
division of the full sample into $18$, into $32$, and into $72$ 
subsamples, respectively.
The results with smaller subsamples have the smaller amplitude.\label{fig:P0s}
}
\end{figure}
%
%
\begin{figure}[h]
\begin{center}
    \includegraphics[width=0.4\textwidth]{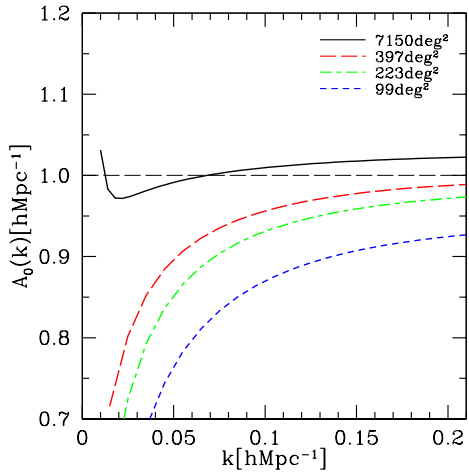}
\end{center}
\caption{$P^{\rm conv}_0(k)/P_0(k)=A_0(k)$ as a function of $k$ for
different divisions of the full sample. From the top to the bottom, 
the curves correspond to the cases with 
no division of the full sample, to the division into $18$, $32$, 
and $72$ subsamples, respectively.
 Here, we assumed the
$\Lambda{\rm CDM}$ cosmology with $\Omega_m=0.28$, $n_s=0.96$,
$\sigma_8=0.8$. The nonlinear power spectrum model, Equation (\ref{PDn}), with 
$\sigma_v={\rm 370km/s}$ is adopted.\label{fig:P0conv}
}
\end{figure}
\begin{figure}[h]
\begin{center}
    \includegraphics[width=0.6\textwidth]{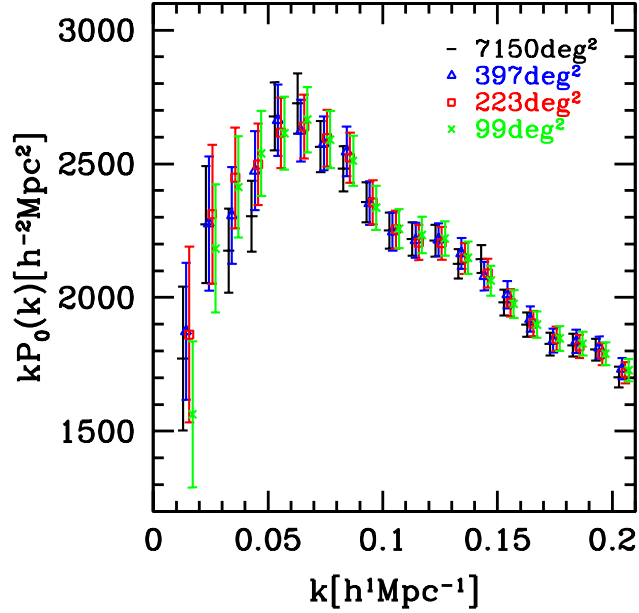}
\end{center}
\caption{ Convolved monopole spectrum divided by the factor $A_0(k)$, 
i.e., $kP^{\rm conv}_0(k)/A_0(k)$. \label{fig:P0corr}}
\end{figure}
\begin{figure}[h]
\begin{center}
    \includegraphics[width=0.9\textwidth]{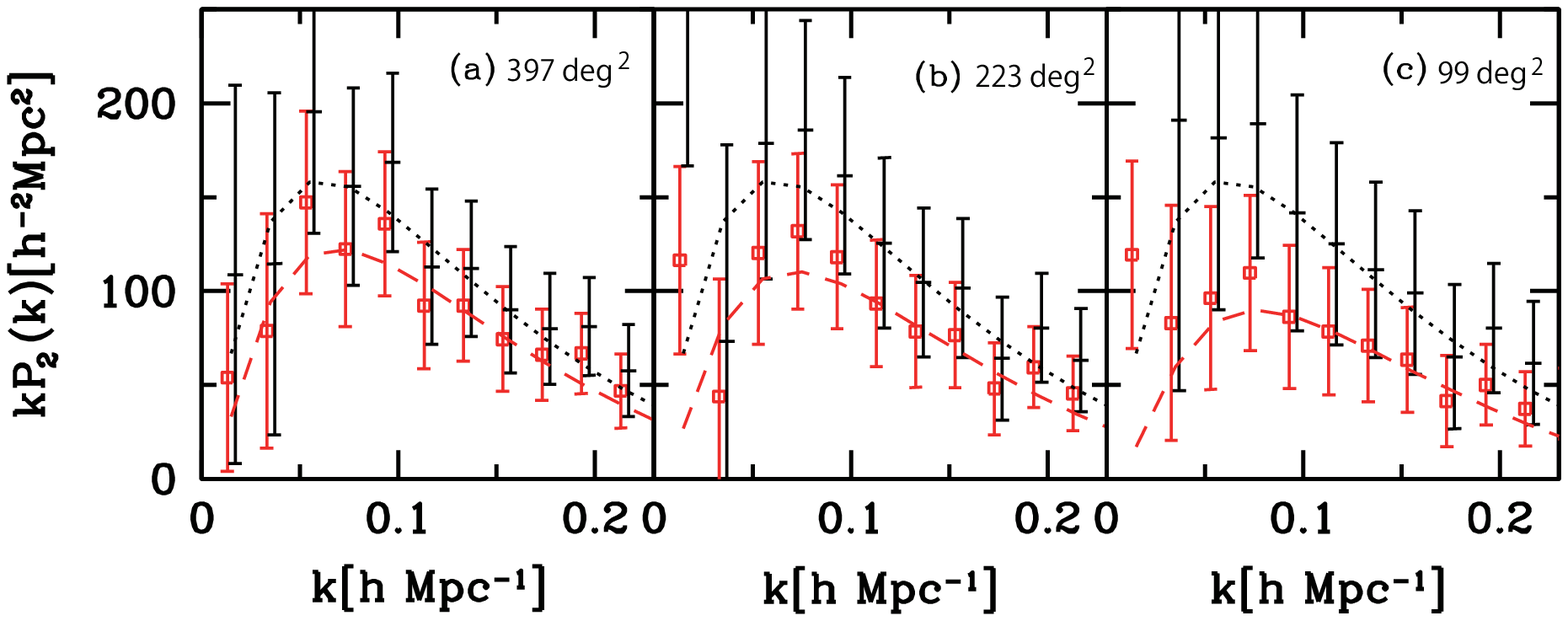}
\end{center}
\caption{Quadrupole power spectra multiplied by the wavenumber $k$
for different divisions of the full sample, with mean 
survey areas ${\rm 397deg^2}$ (a,~left panel), ${\rm 223deg^2}$ 
(b,~center panel),and ${\rm 99deg^2}$ (c,~right panel), respectively. 
The dotted curve is our theoretical model for $kP_2(k)$, 
while the dashed curve is $kP^{\rm conv}_2(k)$ from Equation (\ref{convP2f}), 
where we used the same theoretical model as that in Fig.~\ref{fig:P0conv}. 
The squares with the error bars present
the observed convolved spectrum, $kP^{\rm conv}_2(k)$. 
The dashes with the error bars show $kP^{\rm conv}_2(k)/A_2(k)$.\label{fig:P2}
}
\end{figure}
%



\subsection{Measurement of the convolved power spectrum}
Let us demonstrate the convolved multiple power spectrum 
using the SDSS LRG sample from DR7. 
Figure \ref{fig:P0s} shows the 
monopole spectrum $kP^{\rm conv}_0(k)$ (left panel) and the quadrupole 
spectrum $kP^{\rm conv}_2(k)$ (right panel), respectively. The results
with different divisions of the full sample are plotted. Thus, due to
varying strength of the window convolution, 
the spectrum depends on the particular division of the full sample.
The amplitude of the convolved power spectrum 
in Fig.~\ref{fig:P0s} is smaller for divisions with smaller patch sizes. 
Thus, the window effect is more influential for divisions with 
smaller patches.

Figure \ref{fig:P0conv} plots $A_0(k)$ defined by Equation (\ref{Aell})
for the monopole spectrum $\ell=0$. The curves are obtained by 
computing Equation (\ref{convP0f}) with $P_\ell(k)$ corresponding to a 
spatially flat cold dark matter model with 
cosmological constant and $\Omega_m=0.28$, and with the window function
shown in Fig.~\ref{fig:W0W2}.
For the power spectrum we used the nonlinear model, Equation (\ref{PDn}), given 
in the appendix,
where the transfer function without the baryon oscillations is 
used \cite{BBKS}, 
for simplicity. 
The three curves in Fig.~\ref{fig:P0conv} assume different 
divisions of the full sample, whose patch sizes are shown in the legend.

As shown in Fig.~\ref{fig:P0s}, the amplitude of the convolved 
power spectrum depends on the division and the mean size of the 
subsamples. The amplitude becomes smaller when the mean patch area is reduced. 
Figure \ref{fig:P0corr} shows the 
power spectrum $kP^{\rm conv}_0(k)/A_0(k)$, where $A_0(k)$ is obtained 
by evaluating Equation (\ref{convP0f}), assuming the theoretical model from 
Fig.~\ref{fig:P0conv}.
The amplitude of the power spectra becomes almost the same, 
which means that the amplitude is correctly restored, i.e., 
the window effect is properly treated by Equation (\ref{convP0f}).

Figure \ref{fig:P2} shows $kP^{\rm conv}_2(k)$ for 
different divisions of the full sample, with mean patch sizes
${\rm 397deg^2}$ (a,~left panel), ${\rm 223deg^2}$ (b,~center panel), 
${\rm 99deg^2}$ (c,~right panel), respectively. 
The squares with error bars present $kP^{\rm conv}_2(k)$ from the SDSS LRG 
sample, while the dashes with error bars show $kP^{\rm conv}_2(k)/A_2(k)$,  
where $A_2(k)$ is obtained by computing Equation (\ref{convP2f}) in the same 
way as $A_0(k)$.
The dotted curves give the theoretical model for $kP_2(k)$, where we 
used the same theoretical model as that in Fig.~\ref{fig:P0conv}.
The dashed curves are the corresponding $kP^{\rm conv}_2(k)$,
which are obtained by computing Equation (\ref{convP2f}). The ratio
of the dashed curve to the dotted curve gives $A_2(k)$.

\section{Deconvolved power spectrum}
\subsection{Formulation}
In this section, we briefly review the method for deconvolving the window 
effect in the power spectrum measurement, which was developed in 
Ref. \cite{SATO}. 
Taking the Fourier transform of Equation (\ref{FkFk}), we have
\begin{eqnarray}
&&\hspace{-1cm}
\int d^3{k}
e^{-i\bfk\cdot\bfs}
P^{\rm conv}(\bfk)
={1\over (2\pi)^3}
\left[\int d^3k'e^{-i\bfk'\cdot\bfs}P(\bfk')\right]
\left[\int d^3ke^{-i\bfk\cdot\bfs}W(\bfk)\right].
\end{eqnarray}
The inverse transformation of 
\begin{eqnarray}
&&\hspace{-1cm}
\int d^3k'e^{-i\bfk'\cdot\bfs}P(\bfk')
=(2\pi)^3
{\int d^3{k}e^{-i\bfk\cdot\bfs}P^{\rm conv}(\bfk)
\over
\int d^3k''e^{-i\bfk''\cdot\bfs}W(\bfk'')}
\end{eqnarray}
leads to 
\begin{eqnarray}
&&\hspace{-1cm}P(\bfk)
=
\int d^3{ s} e^{i\bfk\cdot\bfs} 
\frac{
 \int d^3{ k'} e^{-i\bfk'\cdot\bfs}P^{\rm conv}(\bfk')
}{
 \int d^3{k''} e^{-i\bfk''\cdot\bfs} W(\bfk'').
}.
\end{eqnarray}
In the case of a discrete density field of a galaxy catalog, 
we must also take the shotnoise into account. 
The estimators for the convolved power spectrum and the window
function are Equations (\ref{estconvP}) and (\ref{estW}), respectively.
We choose the estimator for the deconvolved power spectrum as
\begin{eqnarray}
&&\hspace{-1cm}
P^{\rm dec}(\bfk)
=
\int d^3{s} e^{i\bfk\cdot\bfs} 
\frac{
U(\bfs)
}{
Y(\bfs)
},
\label{Pkdeconvolved}
\end{eqnarray}
where we defined
\begin{eqnarray}
&&\hspace{-1cm}
U(\bfs)= \int d^3{k'} e^{-i\bfk'\cdot\bfs}
 \Bigl[
  |\int d^3s'F(\bfs')e^{-i\bfk'\cdot\bfs'}|^2
  -(1+\alpha)\int d^3s'\bar n(\bfs')\psi^2(\bfs')
 \Bigr]
\\
&&\hspace{-1cm}
Y(\bfs)= \int d^3{k''} e^{-i\bfk''\cdot\bfs}
 \Bigl[
 \bigl|\int d^3s''\alpha n_s(\bfs'')\psi(\bfs'')e^{-i\bfk''\cdot\bfs''}\bigr|^2
 -\alpha\int d^3s''\bar n(\bfs'')\psi^2(\bfs'')
 \Bigr].
\end{eqnarray}
One can measure the deconvolved multipole power spectra 
from Equation (\ref{Pkdeconvolved}) by
\begin{eqnarray}
&&\hspace{-1cm}
P^{\rm dec}_\ell(k)=\frac{1}{V_k}\int_{V_k}d^3kP^{\rm dec}(\bfk)
{\cal L}_l(\hat{\bf e}\cdot\hat\bfk),
\end{eqnarray}
where $V_k$ is a shell in the Fourier space. 
This deconvolved multipole power spectrum 
can be compared with theoretical predictions directly 
without taking the window effect into account.

\subsection{Comparison between the convolved spectrum 
and the deconvolved spectrum}
\begin{figure}[!ht]
\begin{center}
    \includegraphics[width=0.77\textwidth]{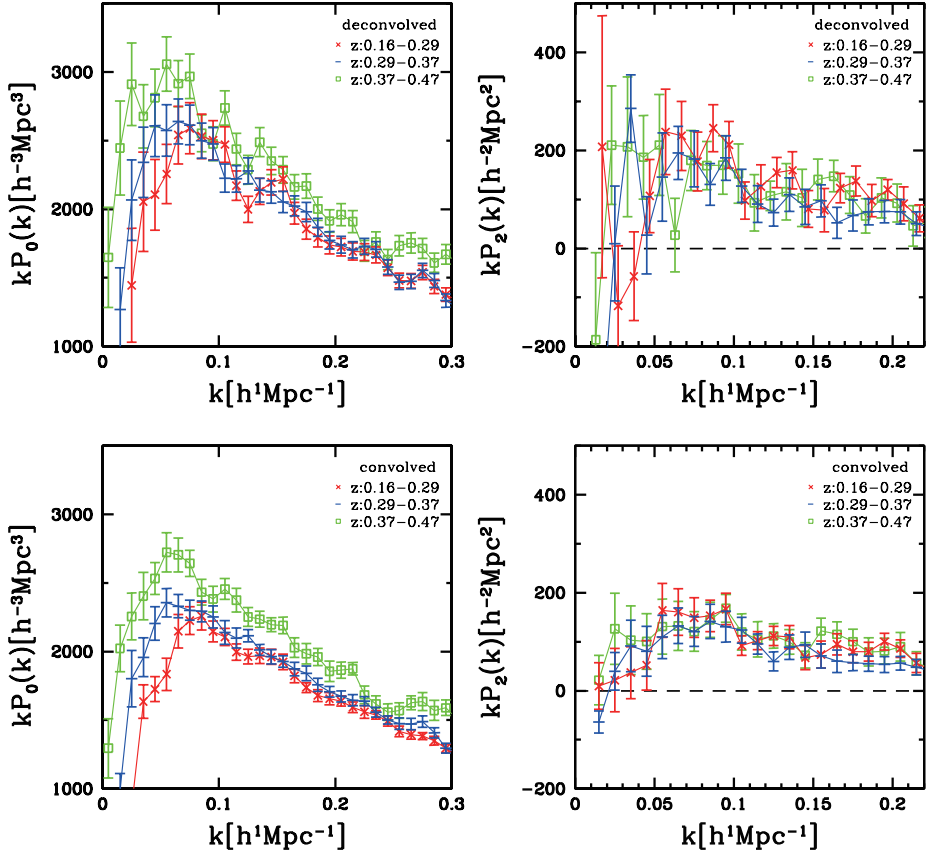}
\end{center}
\caption{Comparison of the multipole power spectra of the 
SDSS LRG sample for redshift ranges $016<z<0.29$ 
(red crosses), 
$0.29<z<0.37$ (blue bars) and $0.37<z<0.47$ (green squares).
The left panels show the monopole spectra,
while the right ones the quadrupole power spectra,
multiplied by the wavenumber.
The upper panels are the deconvolved and 
the lower ones the convolved power spectra.
In this analysis each redshift bin
is divided into 18 angular subsamples whose mean area
is $397$ square degrees.
\label{fig:LRGp0z}
}
\end{figure}

\begin{figure}[!ht]
\begin{center}
    \includegraphics[width=0.72\textwidth]{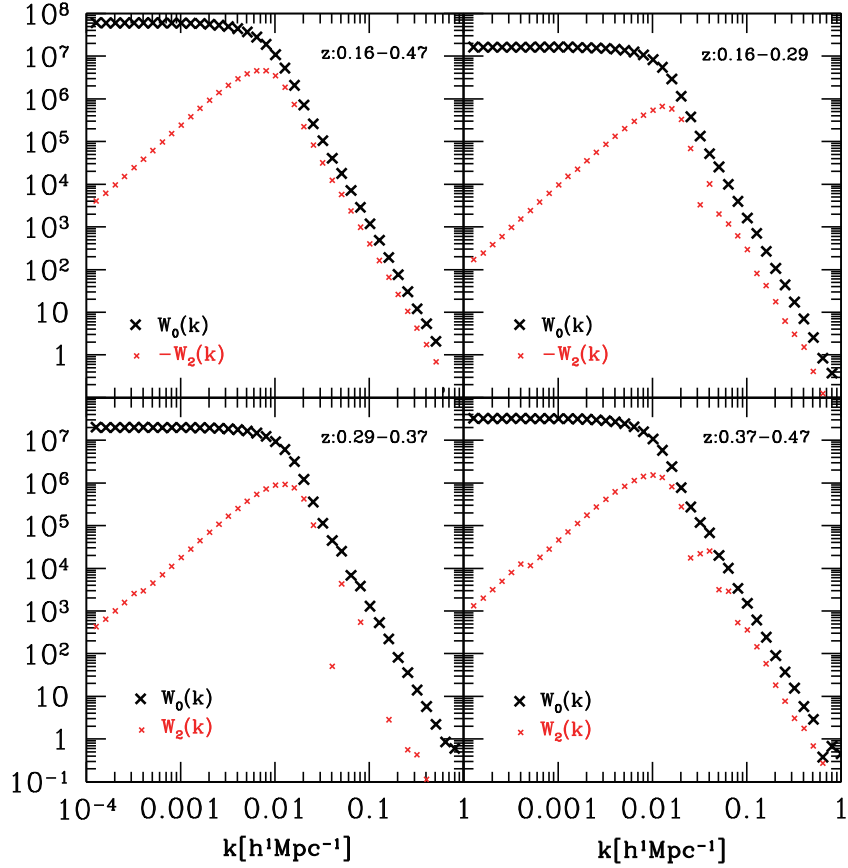}
\end{center}
\caption{The window functions $W_0(k)$ (black crosses) and $|W_2(k)|$
(red small crosses) assuming the full sample is divided into 
subsamples with redshift ranges, $0.16<z<0.29$ (upper right panel), 
$0.29<z<0.37$ (lower left panel) and $0.37<z<0.47$ (lower right panel), 
which correspond to the analysis of Fig.~\ref{fig:LRGp0z}.
The upper left panel is the case using the full redshift range $016<z<0.47$, which 
is the same as the upper right panel of Fig.~\ref{fig:W0W2}. 
\label{fig:LW0W2}
}
\end{figure}

Figure \ref{fig:LRGp0z} compares the convolved and deconvolved 
power spectra of the LRG galaxy sample in the range of 
redshifts $0.16<z<0.29$ (red crosses), $0.29<z<0.37$ 
(blue bars) and $0.37<z<0.47$ (green squares).
The left panels are the monopole spectra,
while the right ones are the quadrupole spectra, multiplied by the wavenumber.
The upper panels are the deconvolved power spectra, while 
the lower ones the convolved spectra, where we
used the division of the full sample into $18$ 
subsamples with mean area $397$ square degrees.
The amplitude of the deconvolved spectrum is larger than 
that of the convolved spectrum.
Figure \ref{fig:LW0W2} shows the multipole moments of the 
window function of the LRG galaxy sample $W_0(k)$ (black large crosses) 
and $|W_2(k)|$ (red small crosses) for the redshift ranges
$0.16<z<0.47$ (upper left panel), $0.16<z<0.29$ (upper right panel),
$0.29<z<0.37$ (lower left panel), and $0.37<z<0.47$ (lower right panel),
respectively. Here each redshift bin
is divided into 18 angular subsamples whose mean area
is $397$ square degrees. The amplitude of $W_0(k)$ in the limit of  
small $k$ is larger when the survey volume is larger. The sign of 
$W_2(k)$ depends on the shape of the subsamples.

\begin{figure}[!ht]
\begin{center}
\begin{tabular}{lr}
    \includegraphics[width=0.26\textwidth,angle=270]{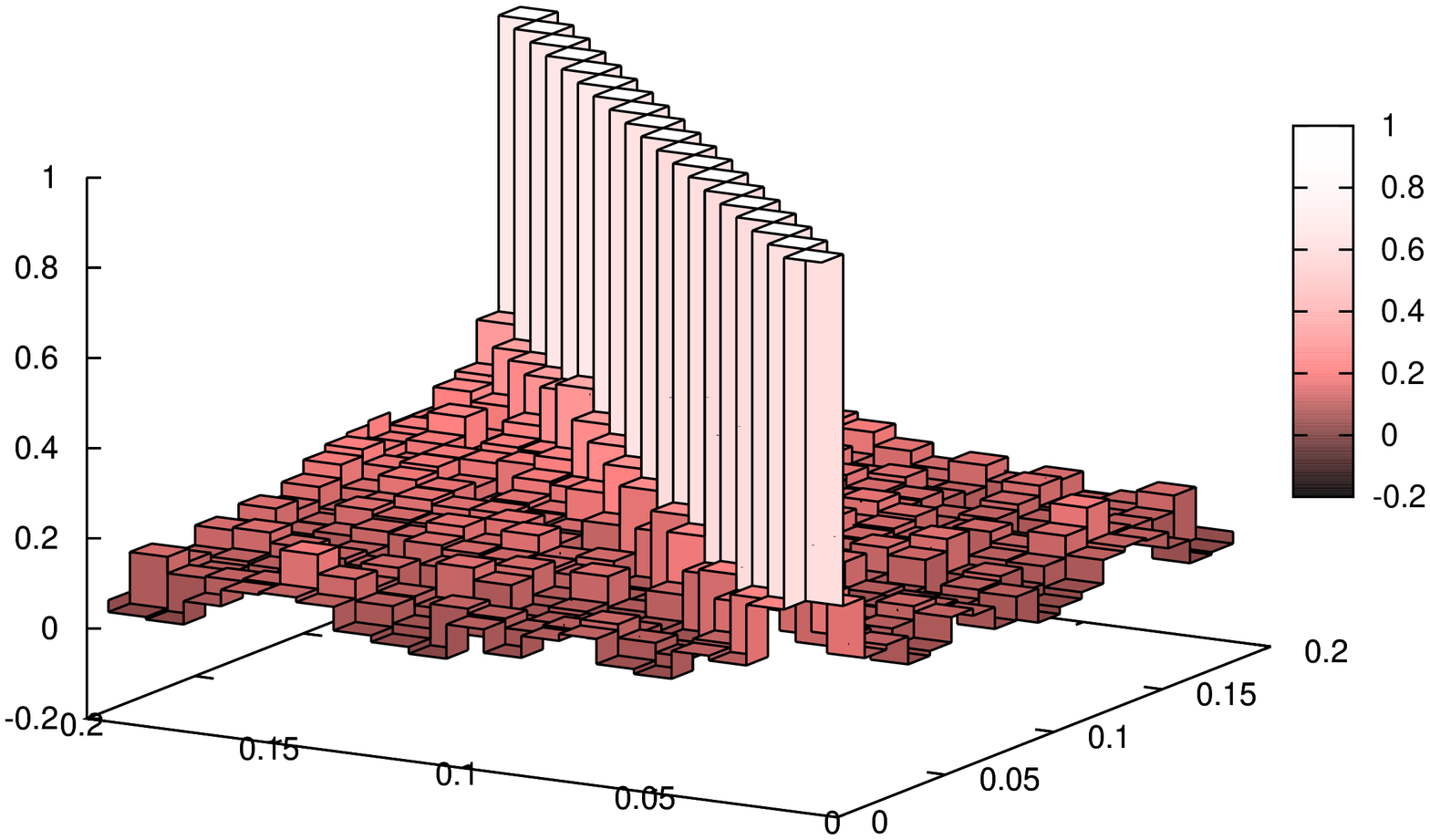}
&   \includegraphics[width=0.26\textwidth,angle=270]{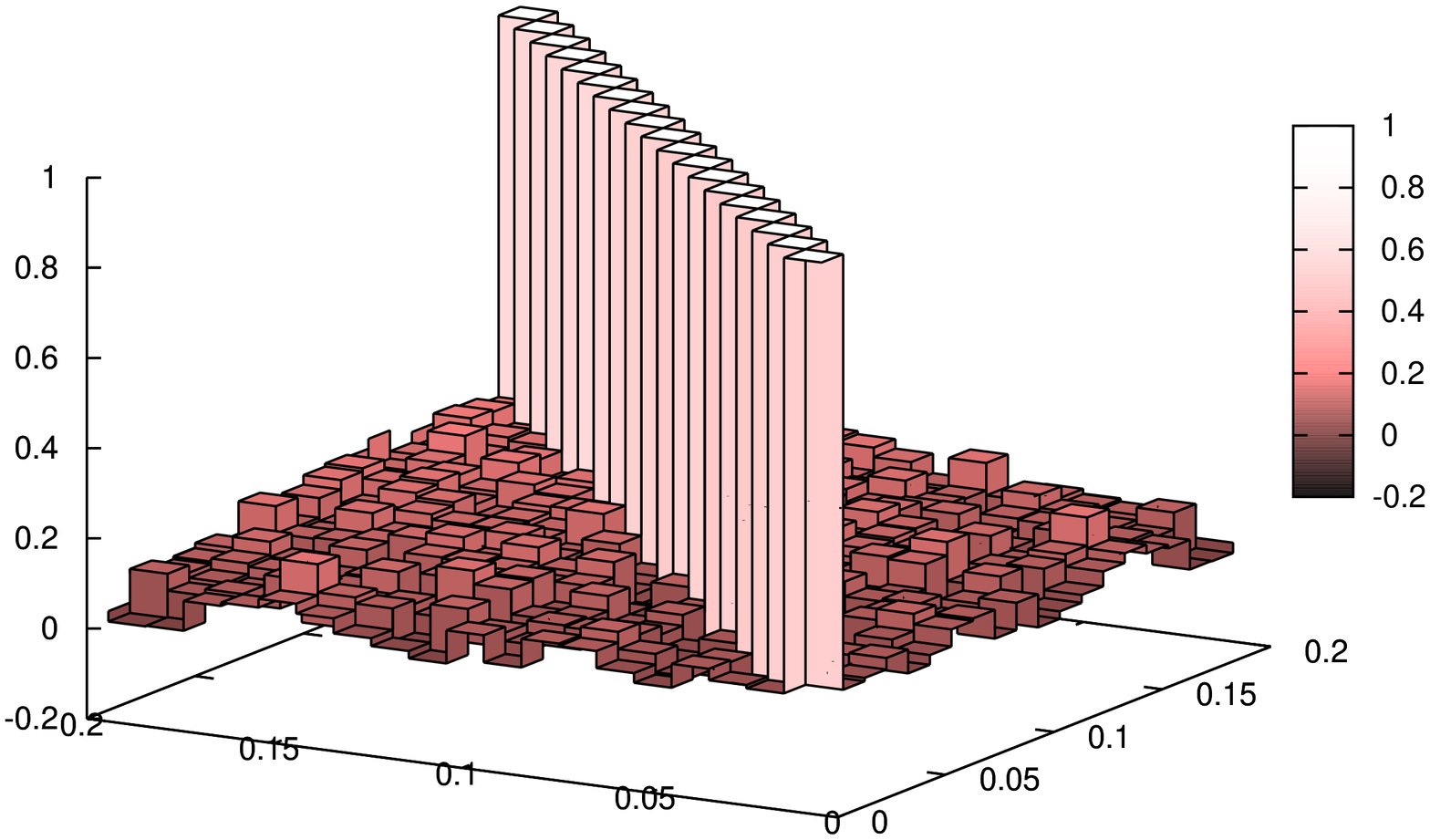}
\\
    \includegraphics[width=0.26\textwidth,angle=270]{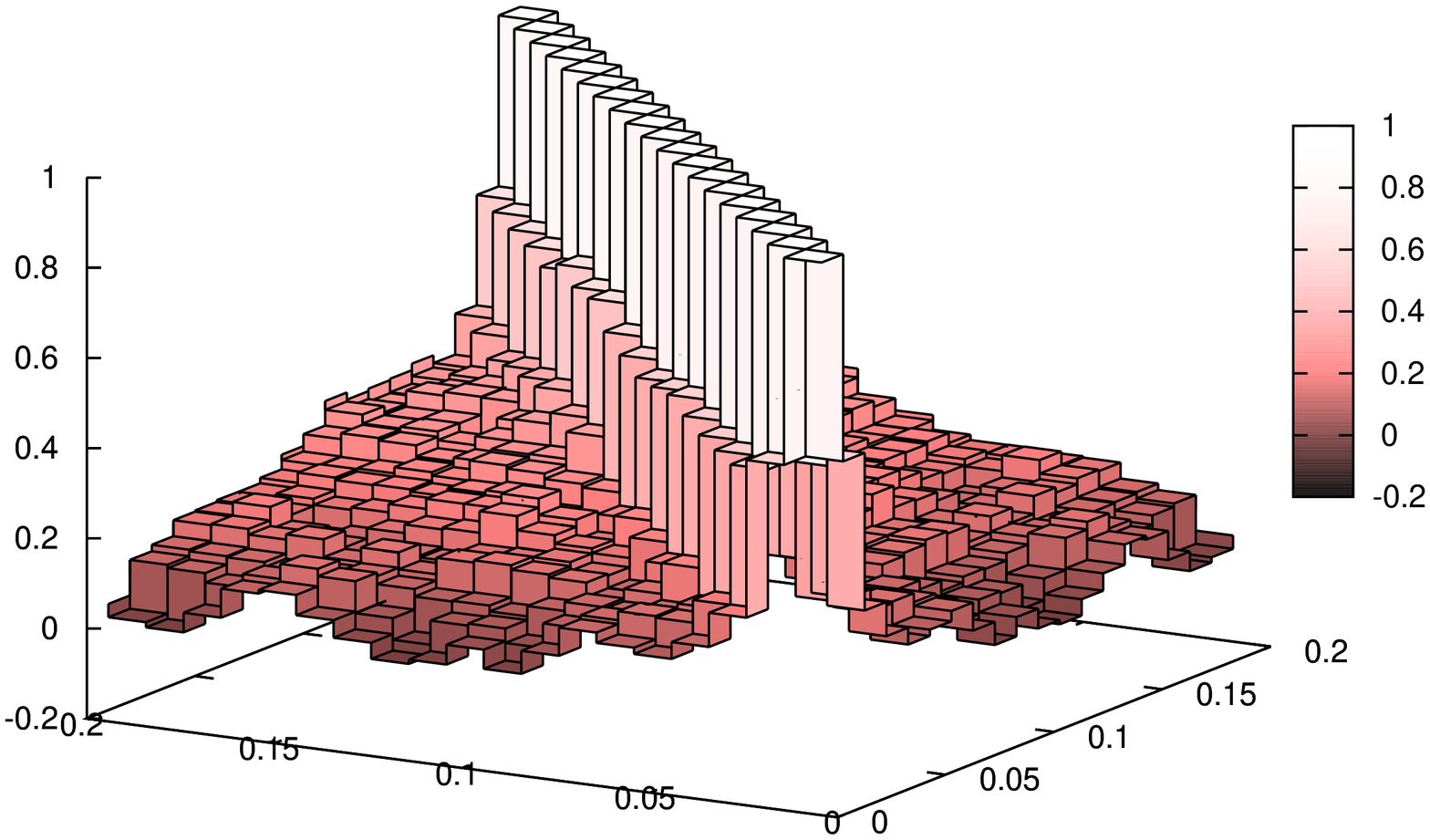}
&   \includegraphics[width=0.26\textwidth,angle=270]{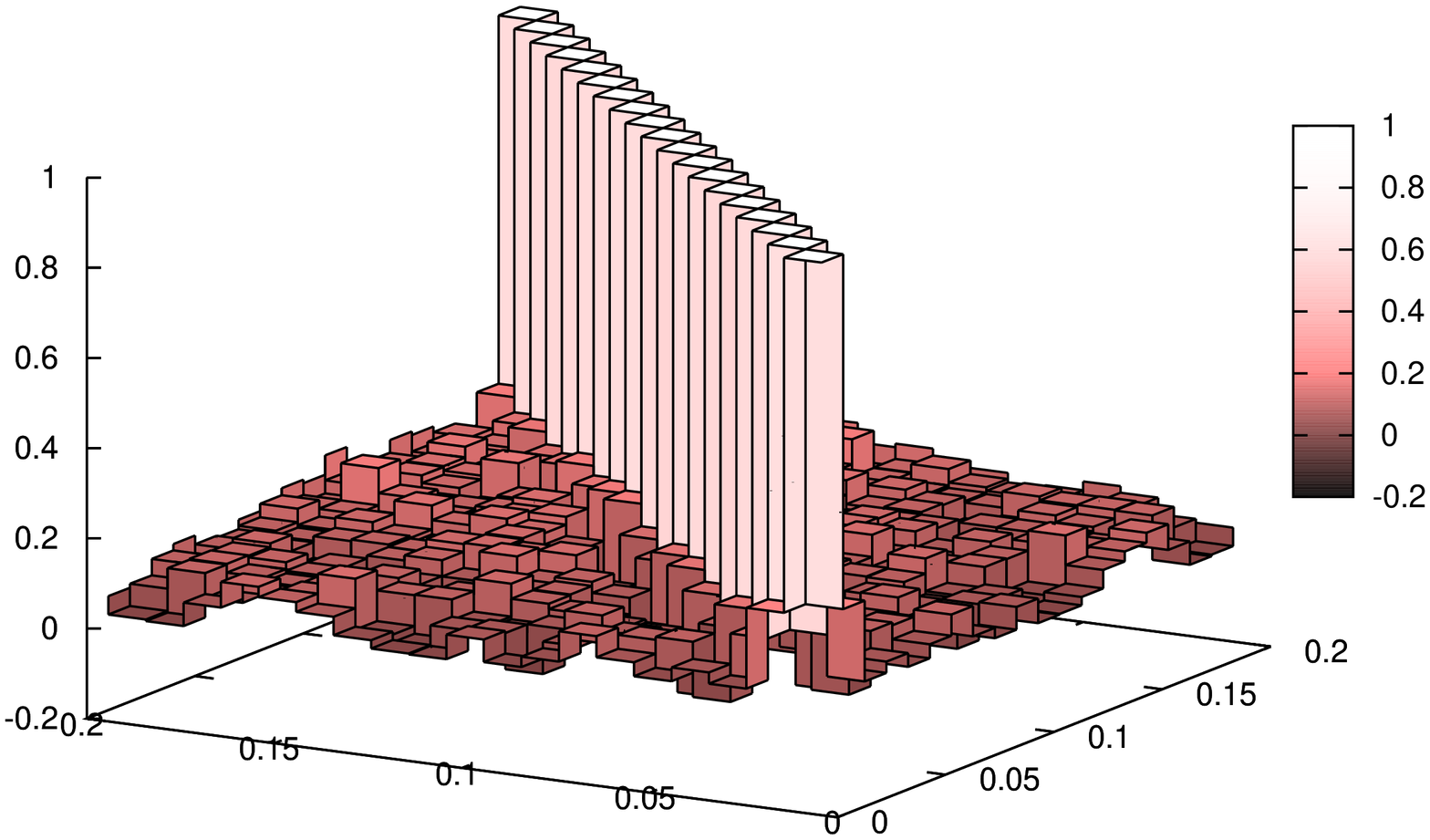}
\\
    \includegraphics[width=0.26\textwidth,angle=270]{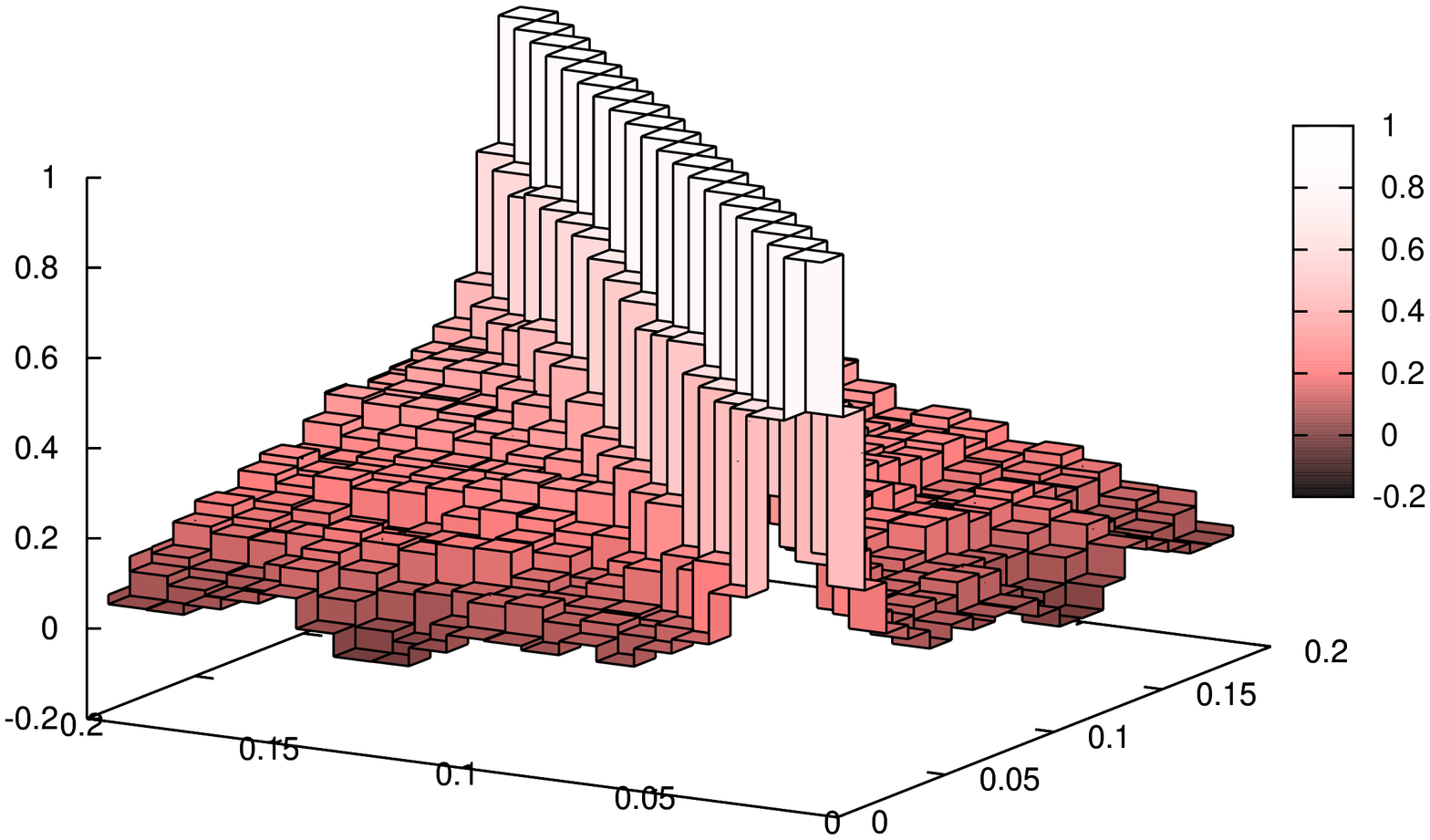}
&   \includegraphics[width=0.26\textwidth,angle=270]{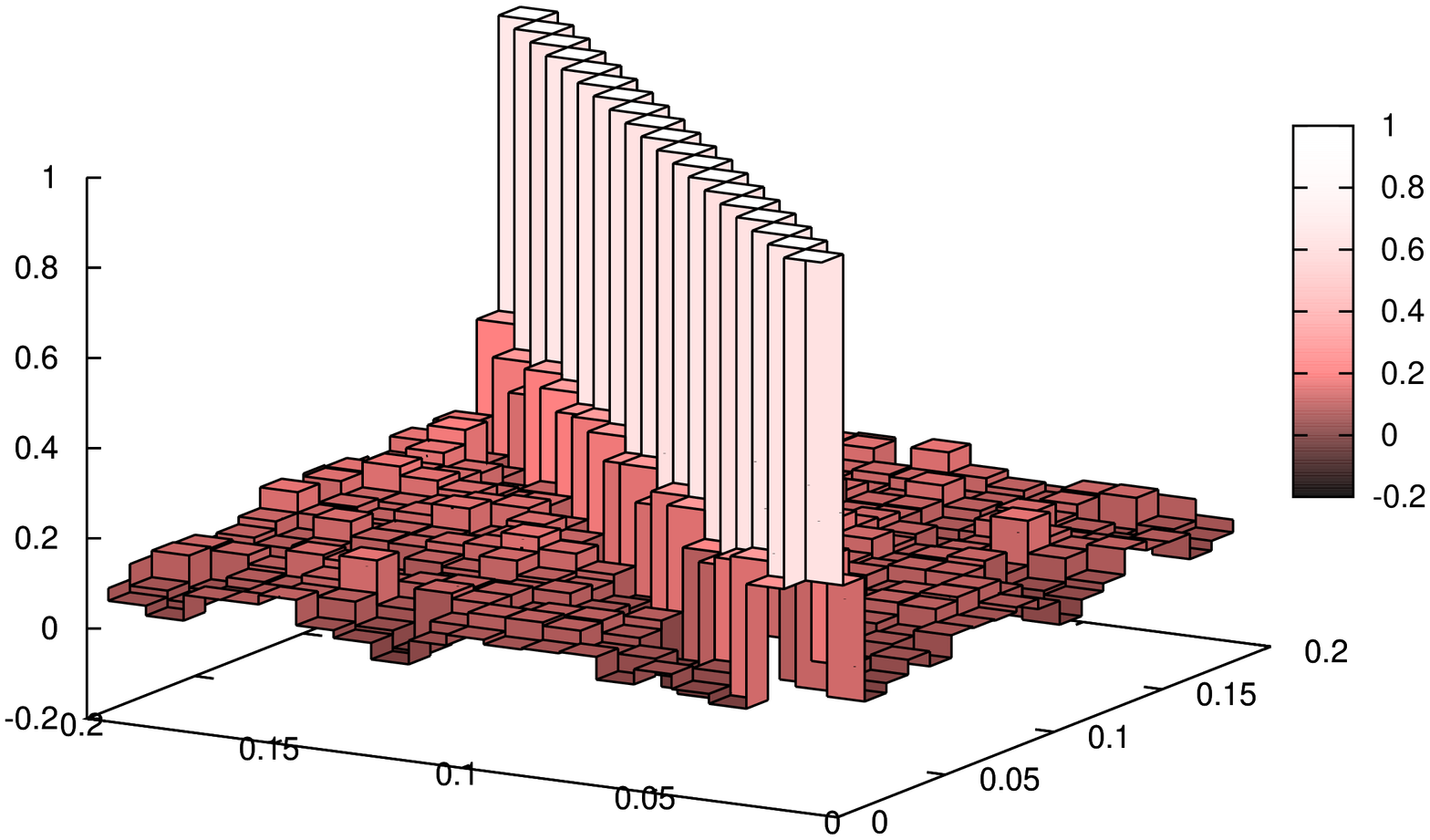}
\\
    \includegraphics[width=0.26\textwidth,angle=270]{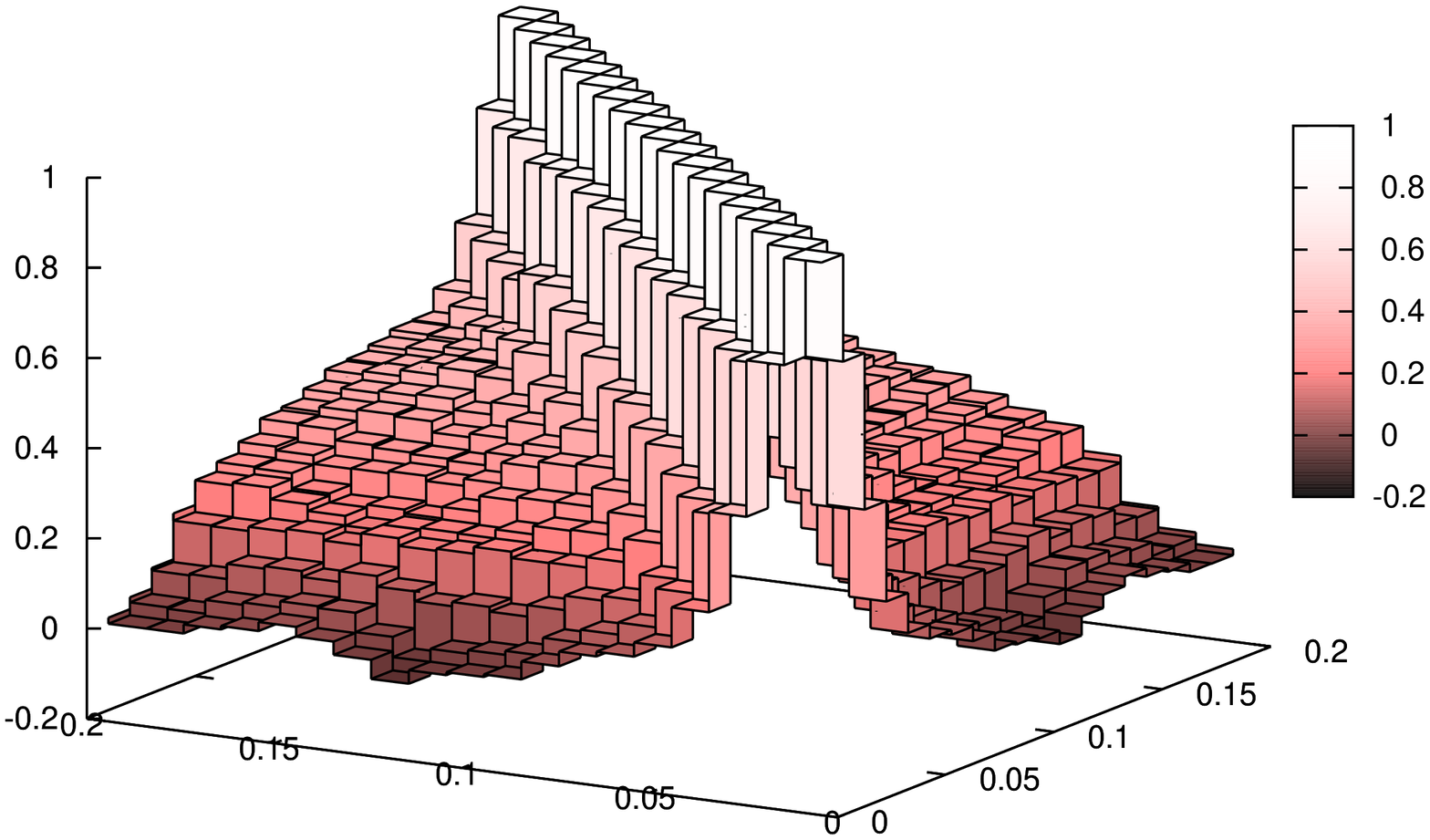}
&   \includegraphics[width=0.26\textwidth,angle=270]{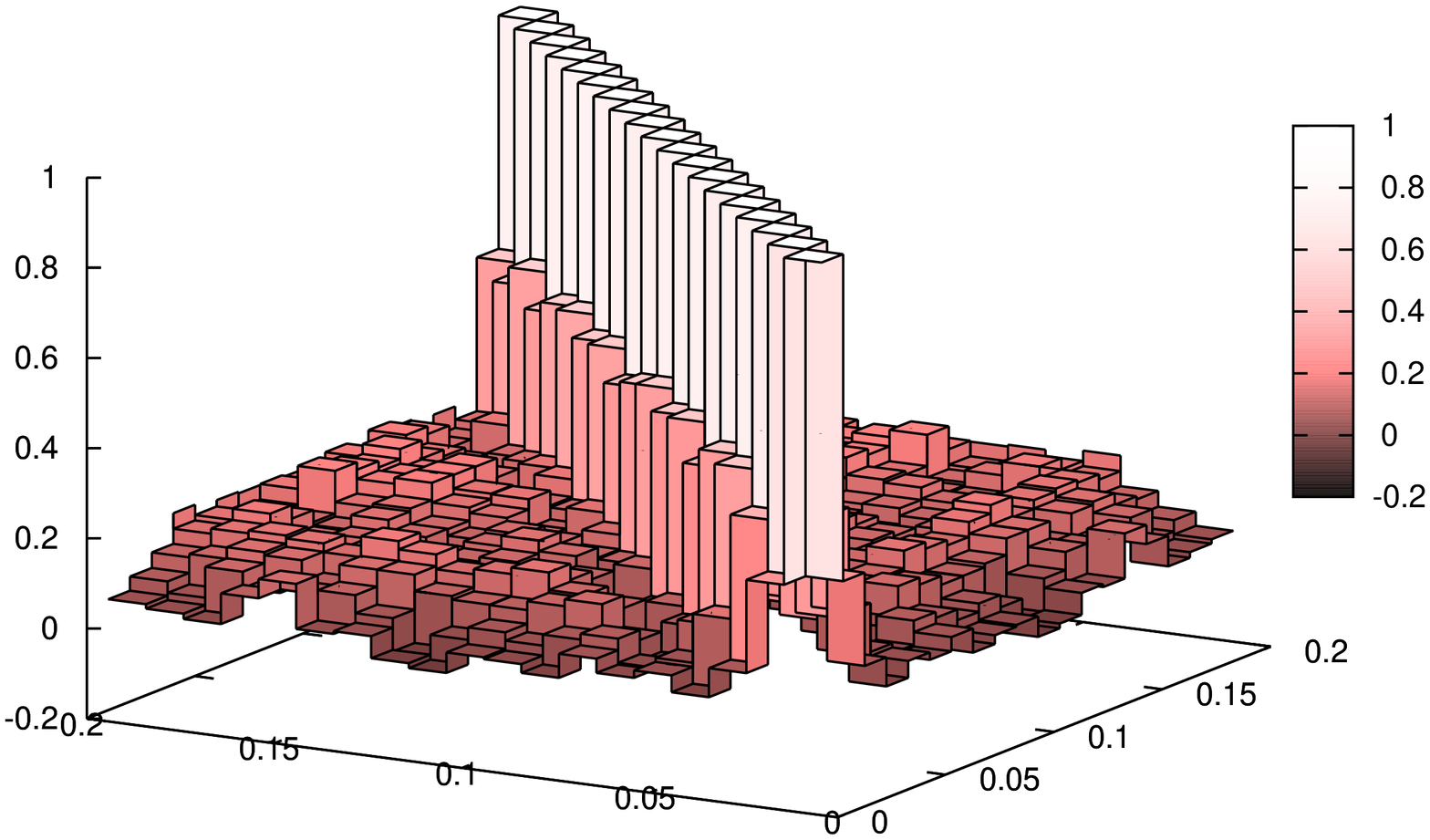}
\end{tabular}
\end{center}
\caption{Correlation matrices of the monopole spectrum, $r_0(k_i,k_j)$,
for the convolved spectrum (left panels) and the deconvolved spectrum 
(right panels), respectively, on the $k_i$ and $k_j$ plane 
($k$ in units of $h{\rm Mpc}^{-1}$). 
The top panels show the case with no division of the full sample, 
while the other lower panels (from bottom to top) 
show the cases when the full sample
is divided into subsamples, whose mean area is $99$, $223$ and $397$
square degrees, respectively.
\label{fig:corrP0}
}
\end{figure}
%

%
\begin{figure}[!ht]
\begin{center}
\begin{tabular}{lr}
    \includegraphics[width=0.26\textwidth,angle=270]{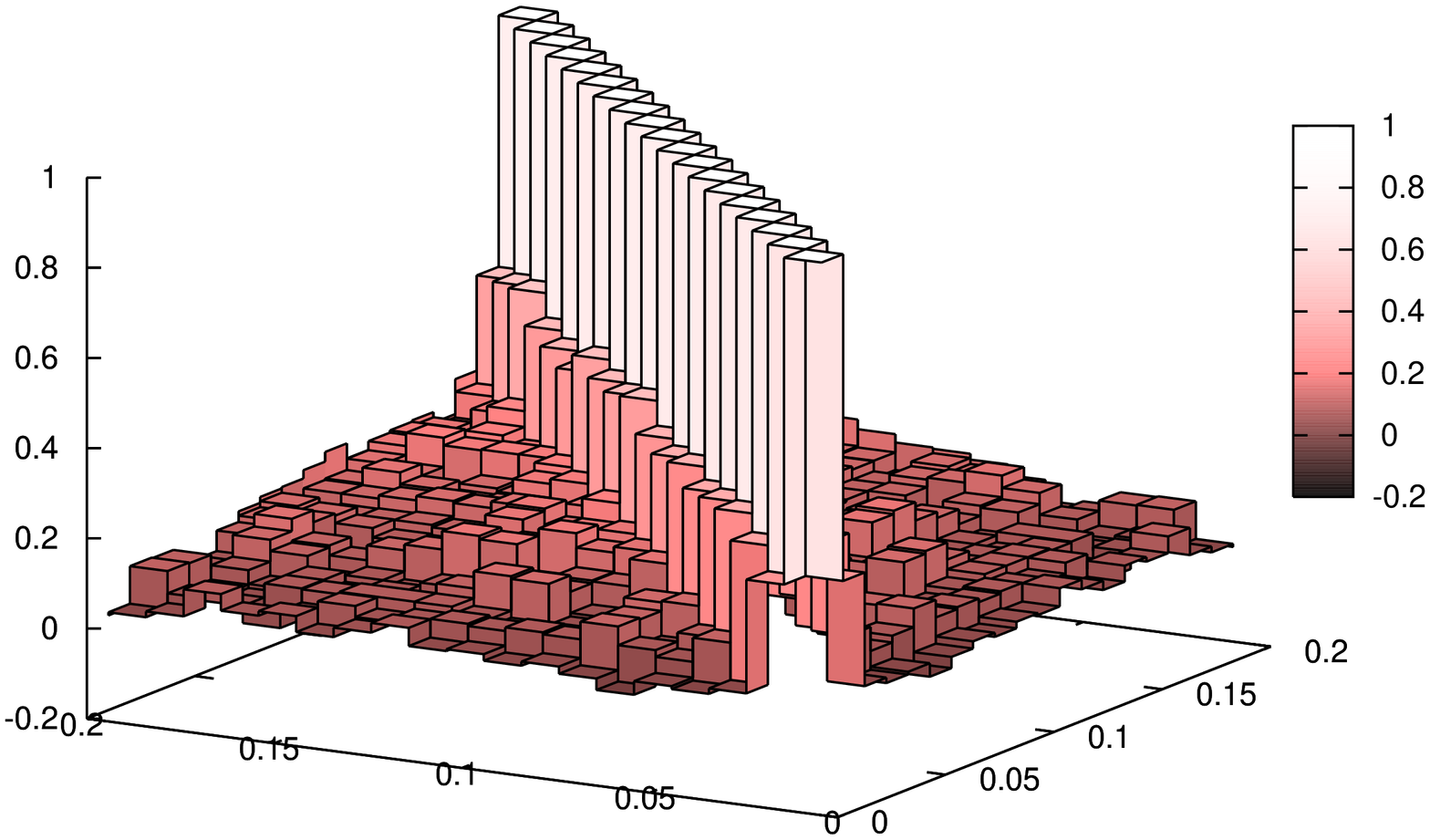}
&   \includegraphics[width=0.26\textwidth,angle=270]{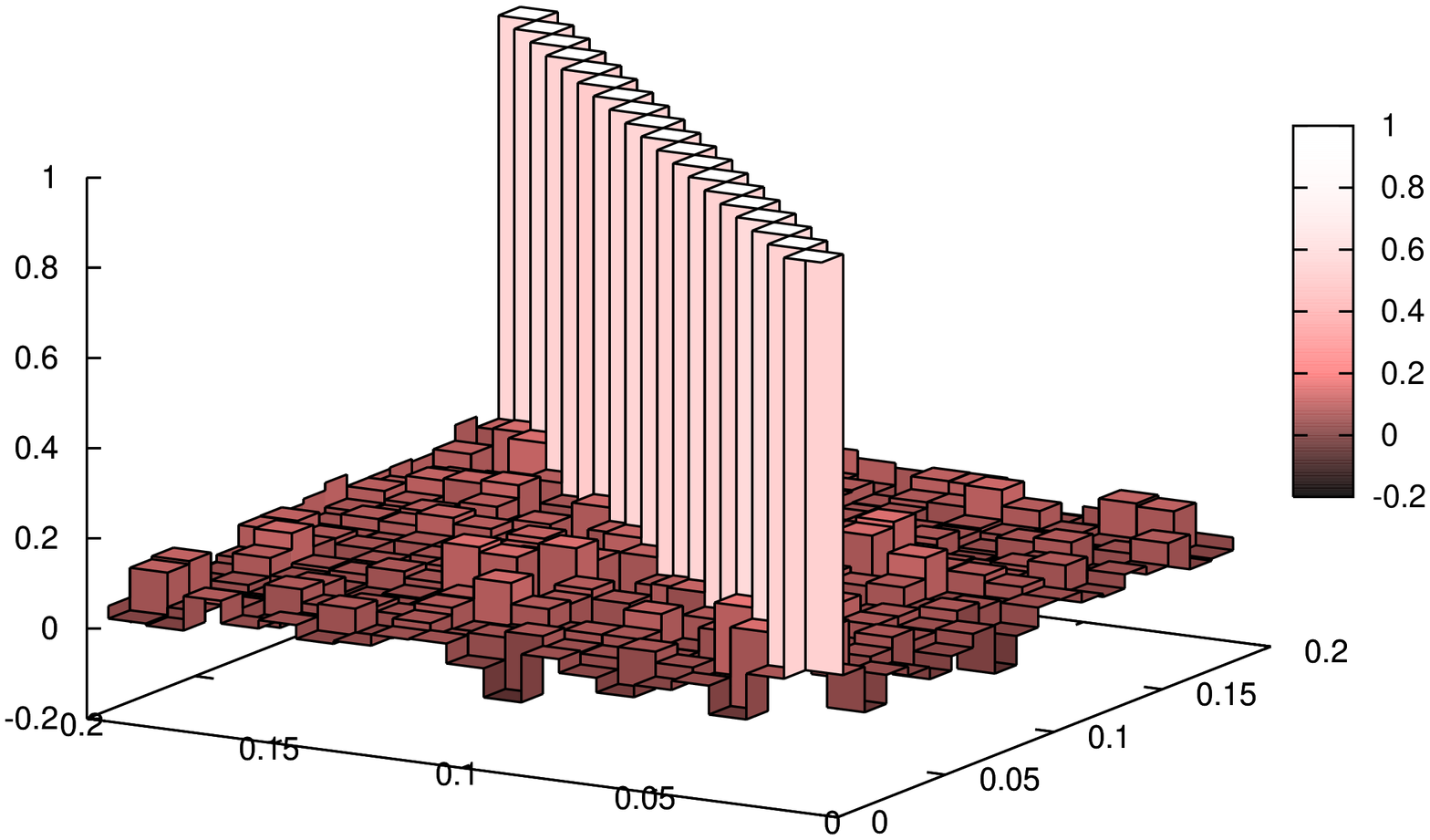}
\\
    \includegraphics[width=0.26\textwidth,angle=270]{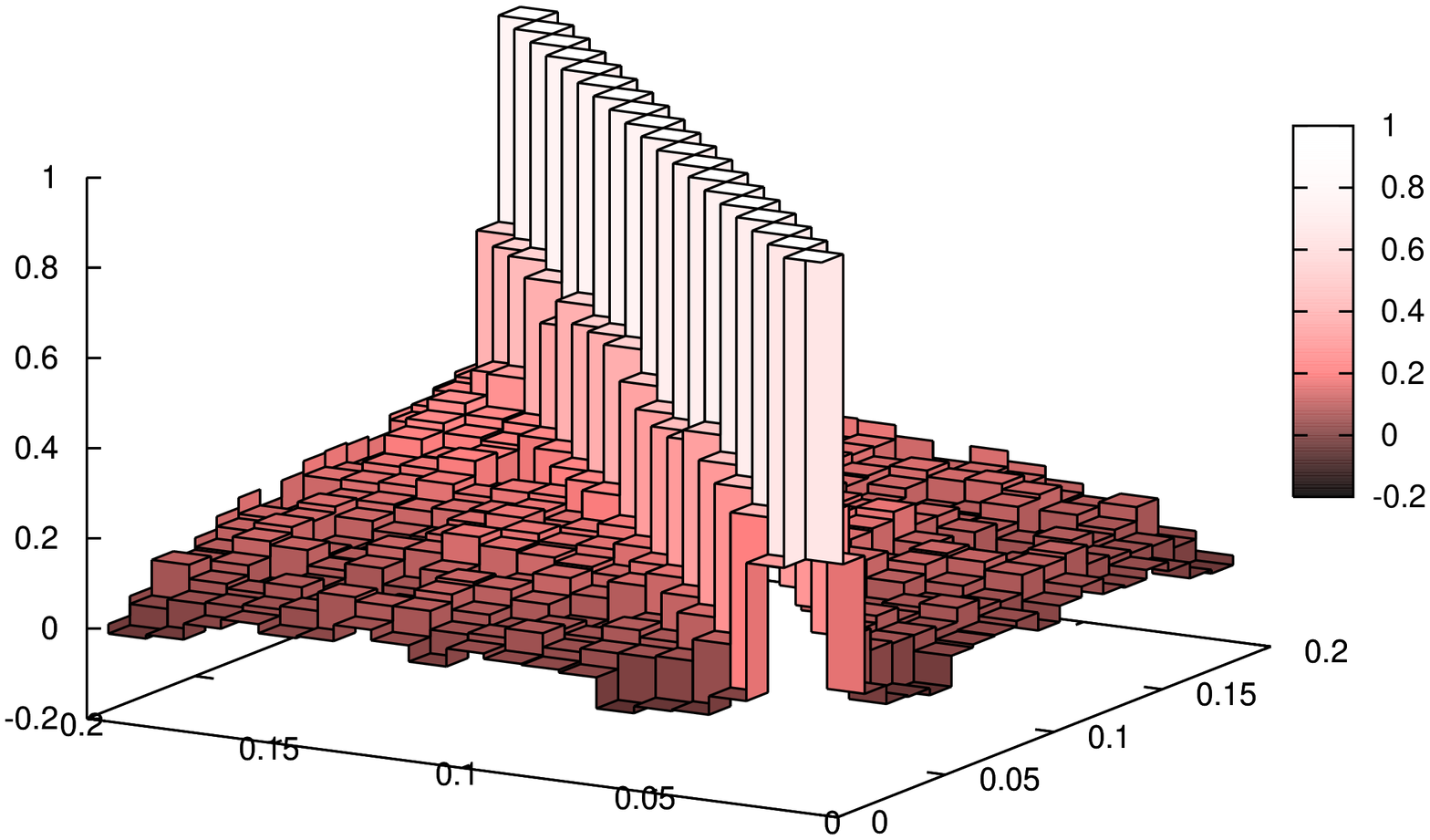}
&   \includegraphics[width=0.26\textwidth,angle=270]{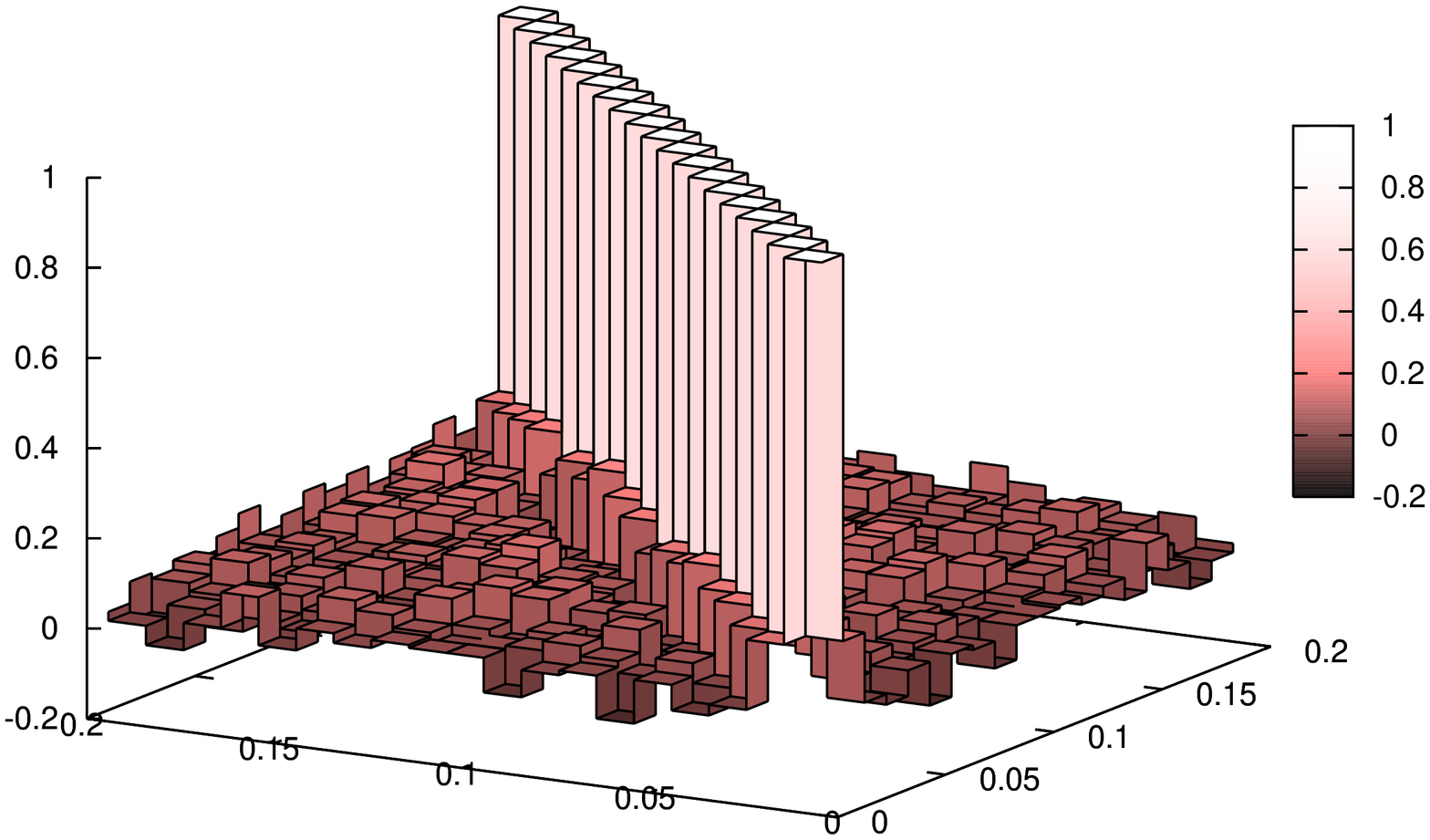}
\\
    \includegraphics[width=0.26\textwidth,angle=270]{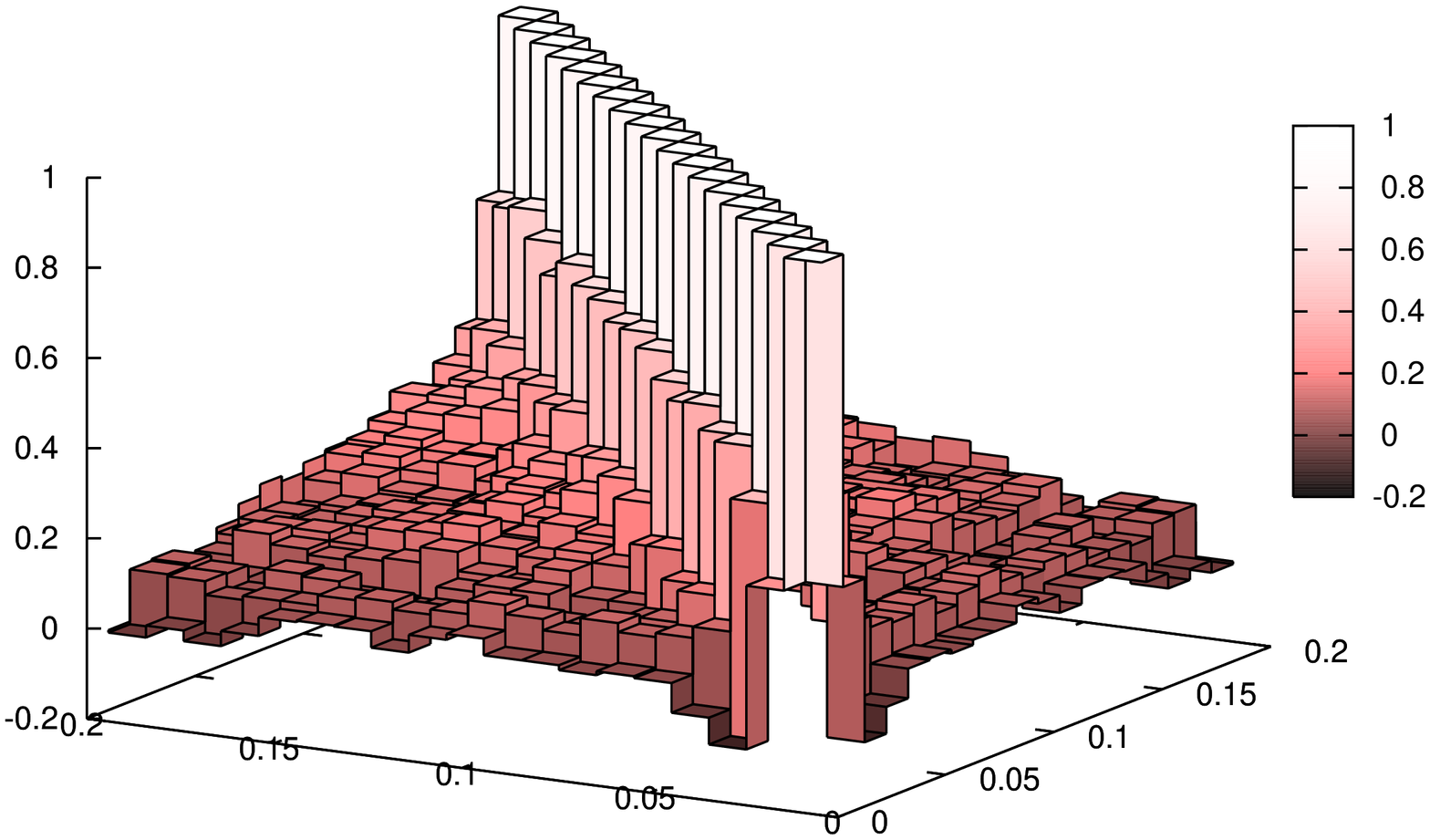}
&   \includegraphics[width=0.26\textwidth,angle=270]{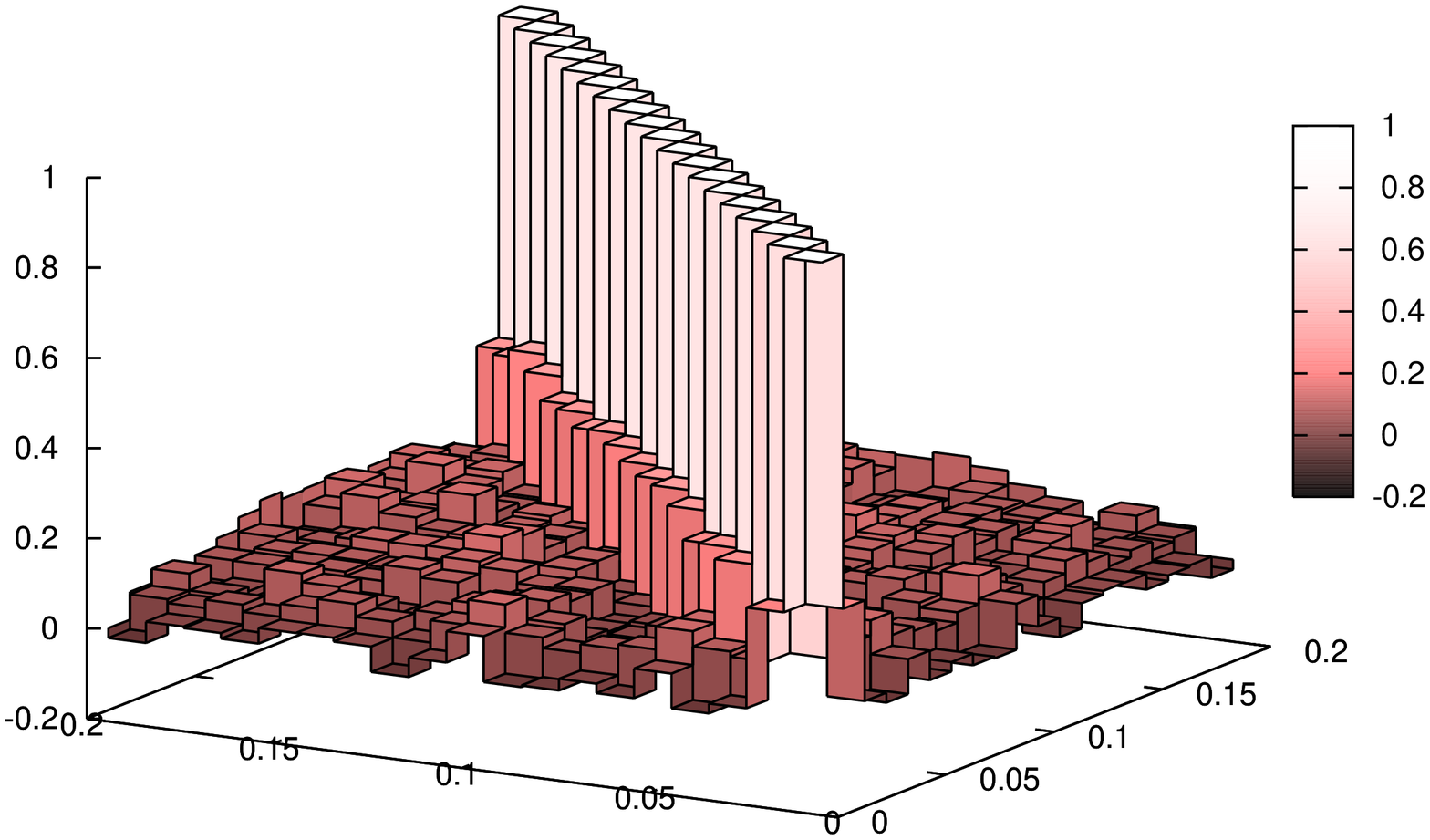}
\end{tabular}
\end{center}
\caption{Correlation matrices of the quadrupole spectrum, $r_2(k_i,k_j)$,
for the convolved spectrum (left panels) and the deconvolved spectrum 
(right panels), respectively. From bottom to top, the panels show the 
cases when the full sample is divided into subsamples, 
whose mean area is $99$, $223$ and 
$397$ square degrees, respectively. 
\label{fig:corrP2}
}
\end{figure}

\subsection{Covariance matrix}
In the following we determine the covariance matrices by utilizing 
mock catalogs corresponding to the SDSS LRG sample.
Our mock catalogs are built by following the procedure 
described in Ref. \cite{Hutsi}.
The covariance matrices for the multipole spectra are defined by 
\begin{eqnarray}
&&\hspace{-1cm}
{C}_{\ell\ell'}(k_i,k_j)
=\Bigl<\Delta P_\ell(k_i)\Delta P_{\ell'}(k_j)\Bigr>
\nonumber
\\
\hspace{-0.5cm}
&\equiv&\Bigl<[P_\ell(k_i)-\left<P_\ell(k_i)\right>]
[P_{\ell'}(k_j)-\left<P_{\ell'}(k_j)\right>]\Bigr>.
\label{covarinceFKP}
\end{eqnarray}
The correlation matrices, which describe the 
correlations between different wavenumbers, are defined by
\begin{eqnarray}
&&\hspace{-1cm} 
r_{\ell}(k_i,k_j)={C_{\ell\ell}(k_i,k_j)\over \sqrt {C_{\ell\ell}(k_i,k_i)
C_{\ell\ell}(k_j,k_j)}}.
\label{correlationmatrix}
\end{eqnarray}
Figure \ref{fig:corrP0} shows the correlation matrices of the monopole
spectrum, $r_0(k_i,k_j)$, on the $k_i$ and $k_j$ plane 
($k$ is in units of $h{\rm Mpc}^{-1}$), which are computed from
$1000$ mock catalogs.
The left panels are for the convolved spectra, while the right ones
for the deconvolved power spectra. 
The top panels show the case without the division of the full sample, 
while the other panels (from bottom to top) represent the cases when 
the full sample is divided into subsamples, whose mean patch sizes are 
$99$, $223$ and $397$ degrees, respectively. 
One can see that the off-diagonal components of the correlation matrices
for the convolved spectrum are larger if the mean area of the subsample 
gets smaller. 
One can also find that the off-diagonal components for the deconvolved 
spectrum get reduced due to the deconvolution. 
The off-diagonal components of the correlation matrix for the
deconvolved spectrum are not completely reduced to zero for 
the cases with the subsamples whose mean patch sizes are small.

Figure \ref{fig:corrP2} shows the correlation matrices for the quadrupole 
spectrum $r_2(k_i,k_j)$. Similar to Fig.~\ref{fig:corrP0}, 
the left panels are for the convolved spectra, while the right panels
for the deconvolved power spectra. 
The division of the full sample into subsamples is needed for 
the quadrupole spectrum. This figure shows the cases when the full 
sample is divided into the subsamples, whose mean areas are $99$, 
$223$ and $397$, respectively, from the bottom to the top panels. 
Similar to the case of the correlation matrix of the monopole spectrum, 
we see that the correlation
between the different wavenumbers becomes significant for the case 
when the full sample is divided into smaller subsamples. 
The effect is more significant when the mean area of the subsample gets 
smaller. The correlation between the different wavenumbers is practically 
de-correlated while using the deconvolved power spectrum. 
Despite of the window deconvolution, the correlation
between the different wavenumbers remains noticeable when the mean area 
of the subsample is small. These features are common to the 
correlation matrices of the monopole.

\section{Summary and Conclusions}
The window effect is very crucial when the power 
spectrum analysis is done by dividing the full sample 
into small subsamples. The division is necessary for obtaining the 
higher order multipole spectra within the distant observer
approximation using the FFT. It is possible to compute the 
higher multipole spectra without the division of the full 
sample \cite{YNKBN}. 
In that case the window effect is not
so significant, however, the FFTs cannot be applied. 
The usage of the FFT is quite useful for performing the Fourier 
transform quickly. Thus, the technique for the treatment of 
the survey window in the power spectrum analysis will be 
quite important.

We investigated the effect of the window function on the multipole 
power spectrum via two different approaches. In the first approach, 
we gave the theoretical formula for the convolved multipole 
power spectra,  Equations (\ref{convP0f}) and (\ref{convP2f}), which 
can be computed by measuring the multipole moments of the window 
function. The multipole moments of the window function were measured 
with the SDSS LRG sample, using the various divisions of the full sample
into subsamples. 
The second approach is the measurement of the power spectrum deconvolved 
from the window effect. The advantage of the deconvolved power spectrum
is the simplicity while comparing with theoretical models. 
The approximate de-correlation between the modes with different wavenumbers
is also the advantage of the deconvolved power spectrum.
We demonstrated the differences between these two approaches to dealing with 
the window effect for the multipole power spectrum.

\noindent

\vspace{2mm}
{\small {\textbf {Acknowledgments.}
This work was supported by Japan Society for Promotion
of Science (JSPS) Grants-in-Aid
for Scientific Research (Nos.~21540270,~21244033).
This work was also supported by JSPS 
Core-to-Core Program ``International Research 
Network for Dark Energy''.
This work was done when T.~Sato and G.~Nakamura 
were PhD students of Graduate School of Physical Sciences, 
Hiroshima University. }}

\def\bf{{}}

\begin{appendix}
\section{Theoretical model for the power spectrum}
In this appendix, we explain the theoretical models adopted in 
the present paper. The simplest model for the galaxy power spectrum 
in the redshift-space is
\begin{eqnarray}
&&P_{\rm gal}(k,\mu)=(b(k)+f\mu^2)^2P_{\rm nl}(k) {\cal D}(\sigma_vk\mu),
\label{PDn}
\end{eqnarray}
where $b(k)$ is the clustering bias, $P_{\rm nl}(k)$ is the nonlinear 
matter power spectrum, ${\cal D}(\sigma_vk\mu)$ is the damping 
factor due to the finger of god effect\cite{Jackson} (see also 
\cite{SargentTurner,TullyFisher}), 
and $\sigma_v$ is the
pair wise velocity dispersion. Assuming an exponential distribution
function for the pairwise velocity, the damping function is 
\begin{eqnarray}
{\cal D}(\sigma_vk\mu)={1\over 1+\sigma_v^2k^2\mu^2/2}.
\end{eqnarray}
For the nonlinear matter power spectrum  $P_{\rm nl}(k)$, 
we adopt the fitting formula by Peacock and Dodds (1994).
\cite{PeacockDodds}.


\end{appendix}

%

%


\end{document}